\documentclass[twoside,twocolumn,9pt]{article}
\usepackage{extsizes}
\usepackage[super,sort&compress,comma]{natbib} 
\usepackage[version=4]{mhchem}
\usepackage[left=1.5cm, right=1.5cm, top=1.785cm, bottom=2.0cm]{geometry}
\usepackage{balance}
\usepackage{mathptmx}
\usepackage{sectsty}
\usepackage{graphicx} 
\usepackage{lastpage}
\usepackage[format=plain,justification=justified,singlelinecheck=false,font={stretch=1.125,small,sf},labelfont=bf,labelsep=space]{caption}
\usepackage{float}
\usepackage{fancyhdr}
\usepackage{fnpos}
\usepackage[english]{babel}
\usepackage{array}
\usepackage{droidsans}
\usepackage{charter}
\usepackage[T1]{fontenc}
\usepackage[usenames,dvipsnames]{xcolor}
\usepackage{setspace}
\usepackage[compact]{titlesec}
\usepackage{hyperref}

\usepackage{booktabs}
\usepackage{url}
\usepackage{miller}

\usepackage{epstopdf}

\definecolor{cream}{RGB}{222,217,201}

\begin{document}

\pagestyle{fancy}
\thispagestyle{plain}
\fancypagestyle{plain}{
\renewcommand{\headrulewidth}{0pt}
}

\makeFNbottom
\makeatletter
\renewcommand\LARGE{\@setfontsize\LARGE{15pt}{17}}
\renewcommand\Large{\@setfontsize\Large{12pt}{14}}
\renewcommand\large{\@setfontsize\large{10pt}{12}}
\renewcommand\footnotesize{\@setfontsize\footnotesize{7pt}{10}}
\makeatother

\renewcommand{\thefootnote}{\fnsymbol{footnote}}
\renewcommand\footnoterule{\vspace*{1pt}%
\color{cream}\hrule width 3.5in height 0.4pt \color{black}\vspace*{5pt}} 
\setcounter{secnumdepth}{5}

\makeatletter 
\renewcommand\@biblabel[1]{#1}            
\renewcommand\@makefntext[1]%
{\noindent\makebox[0pt][r]{\@thefnmark\,}#1}
\makeatother 
\renewcommand{\figurename}{\small{Fig.}~}
\sectionfont{\sffamily\Large}
\subsectionfont{\normalsize}
\subsubsectionfont{\bf}
\setstretch{1.125} 
\setlength{\skip\footins}{0.8cm}
\setlength{\footnotesep}{0.25cm}
\setlength{\jot}{10pt}
\titlespacing*{\section}{0pt}{4pt}{4pt}
\titlespacing*{\subsection}{0pt}{15pt}{1pt}

\fancyfoot{}
\fancyfoot[LO,RE]{\vspace{-7.1pt}\includegraphics[height=9pt]{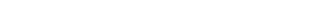}}
\fancyfoot[CO]{\vspace{-7.1pt}\hspace{13.2cm}\includegraphics{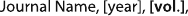}}
\fancyfoot[CE]{\vspace{-7.2pt}\hspace{-14.2cm}\includegraphics{RF}}
\fancyfoot[RO]{\footnotesize{\sffamily{1--\pageref{LastPage} ~\textbar  \hspace{2pt}\thepage}}}
\fancyfoot[LE]{\footnotesize{\sffamily{\thepage~\textbar\hspace{3.45cm} 1--\pageref{LastPage}}}}
\fancyhead{}
\renewcommand{\headrulewidth}{0pt} 
\renewcommand{\footrulewidth}{0pt}
\setlength{\arrayrulewidth}{1pt}
\setlength{\columnsep}{6.5mm}
\setlength\bibsep{1pt}

\makeatletter 
\newlength{\figrulesep} 
\setlength{\figrulesep}{0.5\textfloatsep} 

\newcommand{\topfigrule}{\vspace*{-1pt}%
\noindent{\color{cream}\rule[-\figrulesep]{\columnwidth}{1.5pt}} }

\newcommand{\botfigrule}{\vspace*{-2pt}%
\noindent{\color{cream}\rule[\figrulesep]{\columnwidth}{1.5pt}} }

\newcommand{\dblfigrule}{\vspace*{-1pt}%
\noindent{\color{cream}\rule[-\figrulesep]{\textwidth}{1.5pt}} }

\makeatother

\twocolumn[
  \begin{@twocolumnfalse}
{\includegraphics[height=30pt]{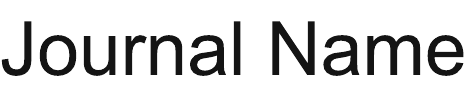}\hfill\raisebox{0pt}[0pt][0pt]{\includegraphics[height=55pt]{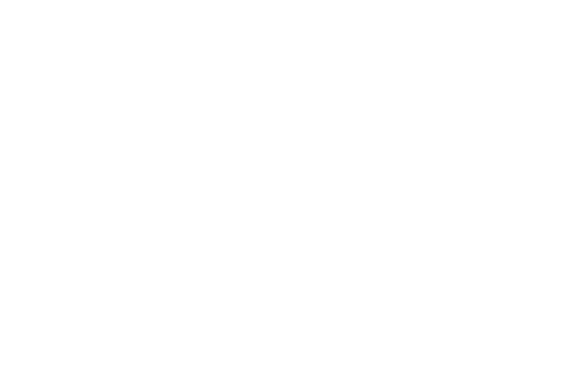}}\\[1ex]
\includegraphics[width=18.5cm]{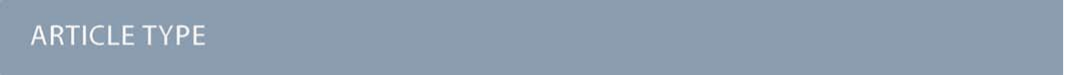}}\par
\vspace{1em}
\sffamily
\begin{tabular}{m{4.5cm} p{13.5cm} }

\includegraphics{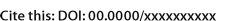} & \noindent\LARGE{\textbf{Machine-learned interatomic potential for titanium carbide \mbox{MXenes:} Application to ion irradiation simulations$^\dag$}} \\
\vspace{0.3cm} & \vspace{0.3cm} \\

 & \noindent\large{Jesper Byggmästar$^{\ast}$\textit{$^{a}$}} \\

\includegraphics{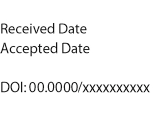} & \noindent\normalsize{A computationally efficient and accurate machine-learned (ML) interatomic potential is developed for bare Ti$_{n+1}$C$_n$ MXenes. With a diverse set of structures computed with density functional theory, the trained ML potential demonstrates good accuracy and robustness to a wide range of bond distances and environments, making it a useful tool for molecular dynamics simulations of MXenes subjected to mechanical load or irradiation. The ML potential is applied to simulations of light and heavy ion irradiation, gathering insight into the statistics and probabilities of sputtering, reflection, defect creation, and implantation into bare Ti$_{n+1}$C$_n$ MXene sheets. The results provide guidelines for defect engineering of MXenes through ion irradiation and implantation. Additionally, the ML potential development provides a landmark recipe for enabling machine-learning-driven atomistic simulations of other MXenes.}

\end{tabular}

 \end{@twocolumnfalse} \vspace{0.6cm}

  ]

\renewcommand*\rmdefault{bch}\normalfont\upshape
\rmfamily
\section*{}
\vspace{-1cm}


\footnotetext{\textit{$^{a}$~Department of Physics, P.O. Box 43, FI-00014 University of Helsinki, Finland; E-mail: jesper.byggmastar@helsinki.fi}}

\footnotetext{\dag~Supplementary Information available: document with supplementary results as well as training data, input, and potential files. See DOI: 00.0000/00000000.}




\section{Introduction}

MXenes have quickly become one of the most intensely studied classes of 2D materials since the first composition was synthesised in 2011~\cite{naguib_twodimensional_2011}. They are represented by the general chemical formula $M_{n+1}X_nT$, where $M$ is a transition metal, $X$ is carbon or nitrogen, $T$ represents a surface termination, and $n$ defines the thickness of the sheet in terms of atomic layers. The first and so far most widely studied MXenes are Ti$_{n+1}$C$_n$, although the rapid interest has expanded the family of MXenes to a vast number of possible compositions of $M$ and $X$ elements, including mixtures of many elements in high-entropy MXenes~\cite{nemani_highentropy_2021}. The attraction of MXenes stem from their many appealing properties together with possibilities for tuning those properties~\cite{gogotsi_future_2023}. Properties such as metallic conductivity and hydrophilicity combined with exceptional mechanical stability and flexibility set MXenes apart from other 2D materials. Applications of MXenes, either suggested or already tested, are found in energy storage, flexible electronics, electromagnetic, shielding, catalysis, biomedical applications, protective coating, among others~\cite{gogotsi_mxenes_2021, gogotsi_future_2023, jiang_two-dimensional_2020, nan_nanoengineering_2021, lin_tinbc_2025}.

In parallel with the rise of MXenes, the field of atomistic modelling has been revolutionised by the rapid development of machine-learning-based methods, in particular ML interatomic potentials~\cite{behler_generalized_2007, bartok_gaussian_2010, shapeev_moment_2016, drautz_atomic_2019}. The development of ML potentials has quickly expanded to cover all classes of classical materials. However, despite the enormous popularity and application potential of MXenes, very few attempts have been made to develop dedicated and accurate ML potentials for MXenes. The only exception known to us is a recent study where a ML potential was developed to simulate oxidation of V$_2$C MXene sheets~\cite{hou_unraveling_2023}. With the lack of reliable interatomic potentials, the exploration of MXenes in molecular dynamics (MD) simulations is limited. Previous studies using MD simulations either employ computationally expensive ab initio MD or the few available ML or classical potentials~\cite{ibragimova_first_2021, berger_raman_2023, thanasarnsurapong_accelerating_2025, oyeniran_firstprinciples_2025, borysiuk_molecular_2015, osti_effect_2016, borysiuk_bending_2018, plummer_bondorder_2022}, some of which were not developed or validated for MXenes. Classical interatomic potentials are limited in accuracy and cumbersome to extend to a wider range of MXene compositions, motivating development of modern ML potentials that do not suffer from these deficiencies.

Ion irradiation and implantation are widely used methods to modify and tune properties of materials by introducing defects~\cite{krasheninnikov_ion_2010}. Defect and nanoscale engineering by irradiation is also a prime example of something that can greatly benefit from insight from atomistic simulations. In other 2D materials such as graphene, irradiation effects have been studied extensively both experimentally and computationally~\cite{krasheninnikov_embedding_2009, komsa_twodimensional_2012, lehtinen_effects_2010, thiruraman_irradiation_2019}. The understanding of MXenes subjected to ion irradiation is still limited to a few pioneering experimental studies~\cite{pazniak_ion_2021, benmoumen_structural_2024}, a recent Monte Carlo simulation~\cite{li_ion_2023}, but completely lacking insight from MD simulations due to the absence of reliable interatomic potentials. Experimentally, Pazniak et al. used 60 keV Mn ion irradiation on \ce{Ti3C2T{$_x$}} MXene thin films~\cite{pazniak_ion_2021}. They concluded that the MXene structure remains stable during irradiation, but with significant damage that allows for increased oxygen functionalization and hence possibilities to tune electronic properties of the MXenes~\cite{pazniak_ion_2021}. Benmoumen et al. in turn used light He ion irradiation on \ce{Ti3C2T{$_x$}} MXene films, revealing further possibilities for irradiation-induced defect engineering by tuning the ion beam fluence~\cite{benmoumen_structural_2024}. Both experimental studies as well as ab initio-based Monte Carlo simulations~\cite{li_ion_2023} show preferential sputtering of Ti, but beyond that the irradiation-induced defects and damage mechanisms remain poorly understood.

This work lays the foundation for development of accurate machine-learning-driven simulations of MXenes by constructing a diverse database of structures and training an accurate ML potential for Ti$_{n+1}$C$_n$ MXenes. The accuracy of the ML potential is demonstrated by computing the elastic response, plastic shear response, vacancy formation energies, phonon dispersion, and quasi-static sputtering drag calculations. The ML potential is used to perform MD simulations of He and Ti irradiation with impact energies spanning four orders of magnitude. From the irradiation simulations we quantify the reflection, implantation, and pass-through probabilities, the sputtering yields of Ti and C atoms, as well as the extent of the produced damage in the MXene sheets. 

\section{Methods}

\subsection{Density functional theory calculations}

DFT calculations for training and testing data were carried out using the \textsc{vasp} code~\cite{kresse_ab_1993,kresse_ab_1994,kresse_efficiency_1996,kresse_efficient_1996} with the PBE GGA exchange-correlation functional~\cite{perdew_generalized_1996} and projector augmented wave potentials~\cite{blochl_projector_1994,kresse_ultrasoft_1999} (default for C, $\texttt{Ti\_sv}$ for Ti from version 54). The plane-wave energy cutoff was 500 eV and the maximum $k$-point spacing was set to 0.15 Å$^{-1}$ in the planar directions and always one $k$-point in the direction normal to the MXene plane. Gaussian smearing with width 0.05 eV was used. A vacuum layer of around 40 Å between periodic MXene sheets was included to minimise periodic interactions. Initially, we considered spin-polarisation to account for nonzero magnetic moments in the MXenes. Spin-polarised calculations of a \ce{Ti2C} MXene sheet results in a ferromagnetic order of the magnetic moments of the Ti layers (0.48 $\mu_\mathrm{B}$) and near-zero magnetic moments for the C layer. The energy of the magnetic \ce{Ti2C} MXene is 33 meV/atom lower than without spin-polarisation and the lattice constant 3.08 Å instead of 3.03 Å (see Supplemental document). While this difference is not negligible, convergence issues were encountered in spin-polarised calculations of MXene training structures with large deformations or atomic displacements, preventing a comprehensive sampling of structures for the training database. For this reason, all DFT calculations (except the reference energies for isolated atoms in vacuum) were done without spin polarisation.

\subsection{Tabulated Gaussian approximation potential}
\label{sec:meth_tabgap}

To produce a computationally efficient ML potential, we choose the tabulated Gaussian approximation potential (tabGAP) method~\cite{byggmastar_modeling_2021, byggmastar_simple_2022}, which is a simple ML potential inspired by classical three-body and embedded atom method potentials~\cite{stillinger_computer_1985, daw_embedded-atom_1984, finnis_simple_1984}. The tabGAP method was initially developed and proven accurate for high-entropy alloys, but has since also been applied to pure metals and semiconductors~\cite{byggmastar_multiscale_2022, zhao_complex_2023}. Apart from being computationally efficient, the benefit of tabGAP is good interpretability due to its low dimensionality compared to other ML potentials~\cite{byggmastar_modeling_2021, byggmastar_simple_2022}. TabGAP is based on GAP~\cite{bartok_gaussian_2010}, which uses sparse Gaussian process regression with a given set of descriptors for the local atomic environment:
\begin{equation}
    E_d = \sum_{i}^{N_d} \delta^2_d \sum_s^{M_d} \alpha_{s} K_d (q_d, q_s),
    \label{eq:GAPd}
\end{equation}
where $d$ represents a descriptor, $N_d$ is the number of descriptor environments, $\delta^2_d$ is an energy prefactor, and $s$ loops through the $M_d$ number of sparse descriptor environments from the training data. $\alpha_s$ are the optimised regression coefficients and $K_d$ is the kernel function that computes a similarity value (from 0 to 1) between a descriptor environment from the training data $q_s$ and the given environment $q_d$. $K_d$ is here the squared exponential kernel for all descriptors. For more details about GAP, see~\cite{bartok_gaussian_2015}. The training was done using the QUIP/GAP code~\cite{bartok_gaussian_2010}. All input parameters and training data are available in an open repository~\cite{ZENODO}.

A tabGAP must be trained with only simple low-dimensional descriptors to allow the energy predictions to be tabulated and evaluated efficiently with 1D and 3D cubic spline interpolations~\cite{byggmastar_modeling_2021,glielmo_efficient_2018,vandermause_--fly_2020}. The combination of descriptors used in the training are a two-body descriptor (the interatomic distance $r_{ij}$), a three-body (the permutation-invariant vector $[(r_{ij}+r_{ik}), (r_{ij}-r_{ik})^2, r_{jk}]$~\cite{bartok_gaussian_2015}), and a scalar embedded atom method (EAM) density ($\rho_i = \sum_j^N \varphi_{ij} (r_{ij})$~\cite{byggmastar_simple_2022}) analogous to classical EAM potentials~\cite{daw_embedded-atom_1984}. The pair density contribution is $\varphi_{ij} (r_{ij}) = (1 - r_{ij} / r_\mathrm{cut})^3$, which smoothly reaches 0 at a cutoff distances $r_\mathrm{cut}$. The cutoff distances are 5.0 Å for the two-body and EAM terms and shorter (4.0 Å) for the three-body term for computational efficiency.

The energy of a system of atoms in tabGAP is the sum of a pre-fitted and fixed repulsive pair potential and the machine-learned tabulated GAP predictions:
\begin{equation}
    E_\mathrm{tot.} = E_\mathrm{rep.} + E_\mathrm{GAP}.
\end{equation}
Hence, the target energy (and corresponding forces) during training is $E_\mathrm{DFT} - E_\mathrm{rep.}$ for a given DFT training structure. The repulsive potential is a Ziegler-Biersack-Littmark (ZBL) screened Coulomb potential~\cite{ziegler_stopping_1985}
\begin{equation}
    E_\mathrm{rep.} = \sum_{i<j}^N \frac{1}{4\pi \varepsilon_0} \frac{Z_i Z_j e^2}{r_{ij}} \phi_{ij} (r_{ij}/a) f_\mathrm{cut} (r_{ij}),
\end{equation}
where $a = 0.46848 / (Z_i^{0.23} + Z_j^{0.23})$ as in the universal ZBL potential. The screening function for an element pair AB is $\phi_\mathrm{AB}(x) = \sum_{i=1}^m \zeta_i e^{\eta_i x}$ with $m=3$. The six parameters $\zeta_i$ and $\eta_i$ are fitted to all-electron DFT data from Ref.~\cite{nordlund_repulsive_1997} for each element pair. A cutoff function $f_\mathrm{cut} (r_{ij})$ is used to drive the repulsion to zero before typical MXene bond lengths~\cite{byggmastar_machine-learning_2019}. The repulsive parts of the tabGAP are shown in the Supplemental document.

After training, the energy predictions from each element pair, triplet, and EAM element are computed and tabulated on 1D and 3D grids and evaluated with cubic spline functions $S$ as
\begin{equation}
\begin{aligned}
     E_\mathrm{tabGAP} =
    & \sum_{i < j}^N S_{\mathrm{rep. + 2b}}^\mathrm{1D} (r_{ij}) + \sum_{i, j<k}^N S_\mathrm{3b}^\mathrm{3D} (r_{ij}, r_{ik}, \cos \theta_{ijk}) \\
    & + \sum_{i}^N S_\mathrm{EAM}^\mathrm{1D} \left(\sum_j^N S_\varphi^\mathrm{1D} (r_{ij}) \right).
\end{aligned}
\end{equation}
The tabGAP can be used with the \textsc{lammps} implementation from \url{https://gitlab.com/jezper/tabgap}. The potential files are available from~\cite{ZENODO}.

\subsection{Helium interatomic potentials}

All He interactions are treated by simple pair potentials and added to the tabGAP as tabulated pair energies. For He--He interactions we use the Beck potential~\cite{beck_new_1968} joined to the ZBL potential~\cite{morishita_thermal_2003,juslin_interatomic_2013} (although no He--He interactions are actually present in any of our simulations). For He--C and He--Ti we use the purely repulsive screened Coulomb potentials from Ref.~\cite{nordlund_repulsive_2025a}.

\subsection{Molecular statics and dynamics simulations}

All molecular statics and MD simulations are done with the \textsc{lammps} code~\cite{thompson_lammps_2022}. The energy versus lattice constant curves are calculated by minimizing atomic positions at all strains. The shearing calculations are done by shearing the given layer along the chosen directions and relaxing the atomic positions at every shear strain. The phonon calculations are done with Phonopy~\cite{togo_firstprinciples_2023, togo_implementation_2023} connected to LAMMPS and VASP. The supercell size was $4\times4$ unit cells. The static drag calculations are done statically with no relaxation. Vacancy formation energies are computed in sheets of $4\times4$ unit cells (47, 79, 111 atoms for MXene thicknesses $n=1$, 2, 3) in both DFT and LAMMPS. The correct lattice constant is used for each thickness and method and only atomic positions are relaxed during minimisation. The vacancy formation energy is then computed as $E_\mathrm{f} = E_{\mathrm{vac}-X} - E_\mathrm{pristine} + \mu_X$, where $E_{\mathrm{vac}-X}$ is the total energy of the relaxed $X$ (Ti, C) vacancy MXene structure, $E_\mathrm{pristine}$ is the total energy of the pristine MXene, and $\mu_X$ is the chemical potential of element $X$. For the chemical potential we use the energy of isolated atoms in vacuum, as in~\cite{plummer_bondorder_2022}.

The irradiation simulations were performed at 300 K using He and Ti ions with kinetic energies from 15 eV to 4 keV for He and 15 eV to 100 keV for Ti ions. Before the irradiation, single-layer \ce{Ti2C} and \ce{Ti3C2} sheets of 50--60 Å wide (1200 atoms for \ce{Ti2C} and 2000 atoms for \ce{Ti3C2}) were relaxed at 300 K with the correct lattice constant. The He or Ti ion was directed downwards normal to the surface from a height 10 Å above the top atomic layer. The in-plane coordinates of the incoming ion was randomised to sample impacts all over the MXene surface. For each ion energy, 10000 independent single-impact simulations were run to collect statistics. Each simulation was run for 2 ps, which was enough for sputtering and damage to occur, although not enough to fully relax the damage in the most extreme cases. An adaptive time step was used to accurately simulate high-energy and close-range nuclear collisions The irradiation simulations were done in the $NVE$ ensemble except for a 4 Å thick region $NVT$ around the periodic border for dissipating heat. Energy losses due to electronic stopping may be important at the highest ion energies (100 keV), but is nontrivial to consider in 2D material irradiation and hence neglected~\cite{liebsch_quantitative_2026}. 

After each impact simulation, possibly sputtered Ti and C atoms were counted and recorded. Atoms can sputter downward or upward. The fate of the incoming ion (passed through, reflected, implanted) was also automatically detected and recorded. Coordination defects were analysed based on the C sublattice only, since for both MXenes the C layers have same coordination (which is not the case for the three Ti layers in \ce{Ti3C2}). Coordination numbers of C atoms were computed with a cutoff distance of 2.55 Å. Coordination defects are counted as number of C atoms with coordination numbers differing from that in a perfect MXene sheet (6).

\section{Results and discussion}

\subsection{Training database and ML potential}

\begin{figure*}
    \centering
    \includegraphics[width=0.9\linewidth]{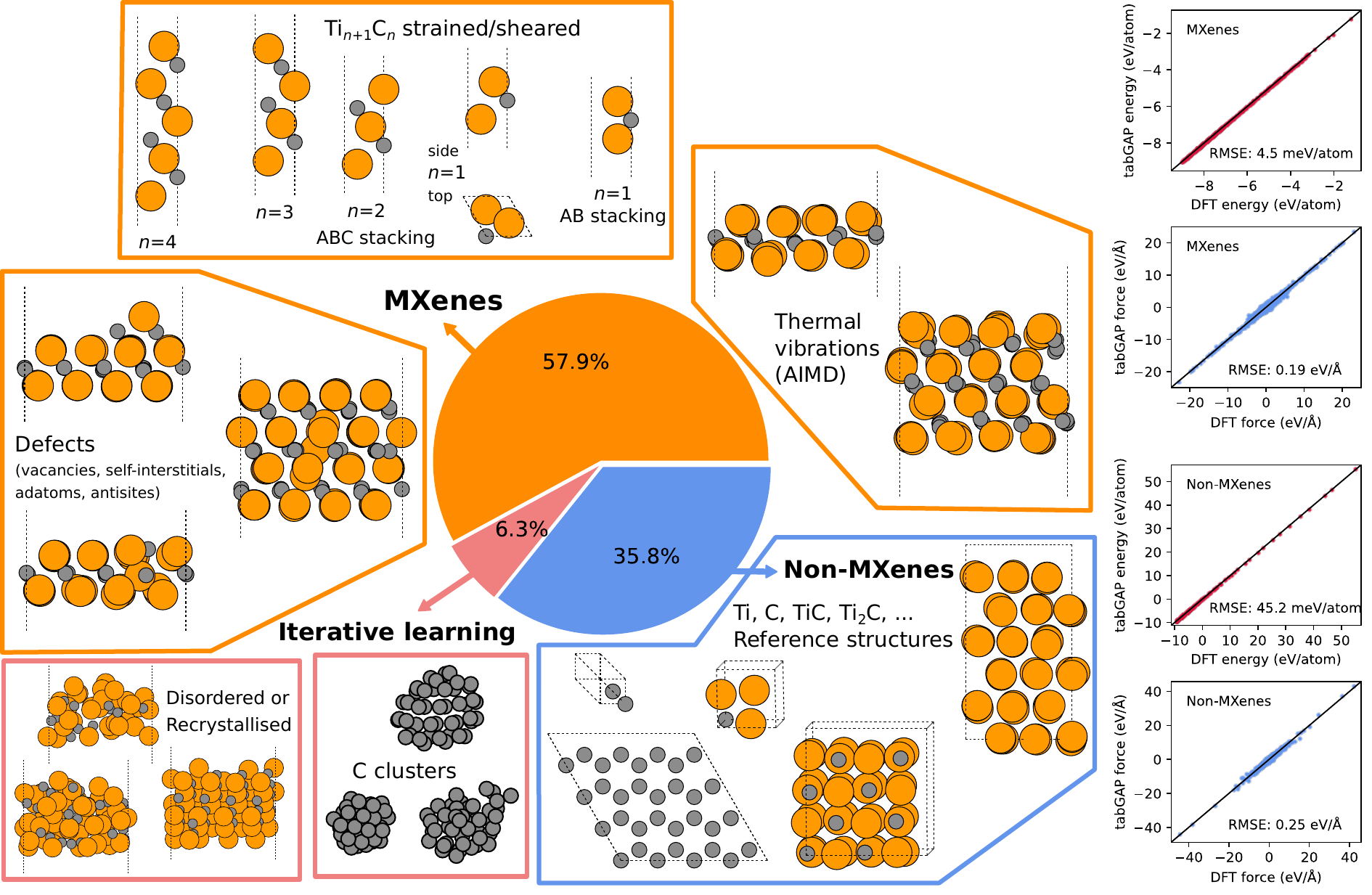}
    \caption{Illustration of the structures constructed and included in the training database for the ML potential, grouped into three classes: MXenes, non-MXenes, and ``iterative learning'' (see text for details). The percentages are computed by counting atoms in all structures of each class (in total 24150 atoms). MXenes include both ABC and AB stacking and from single-layer up to Ti$_5$C$_4$. Non-MXenes include a small set of bulk Ti, C, and TiC structures as well as graphene to ensure correct thermodynamic stability of the MXenes. The right panel shows train errors in parity plots of the ML potential energy and force compared to the DFT reference values.}
    \label{fig:db}
\end{figure*}

The key to develop a robust and accurate ML potential is a training set with structures that cover all relevant bonding environments expected in the applied simulations. This is crucial for avoiding extrapolation, which ML models are notoriously bad at. We chose to gradually construct the training database by a combination of manual creation of structures, sampling from ab initio MD (AIMD) simulations, and iterative training using ML potentials trained to the gradually growing training database. An illustrative summary of the training structures is shown in Figure~\ref{fig:db}. The final training set includes 1522 structures containing in total 24150 atoms.

The initial part of the training database consisted of single unit cells of Ti$_{n+1}$C$_n$ MXene sheets of ABC and AB stacking. Wide ranges of lattice spacings were sampled by straining and shearing in the planar directions, capturing the elastic and plastic response of the MXenes. Thermal and vibrational properties were sampled by picking frames from AIMD simulations of Ti$_2$C and Ti$_4$C$_3$ MXene sheets at 300 and 1000 K. Additionally, we also constructed a set of non-MXene structures consisting of bulk TiC and other Ti-C structures from the Materials Project~\cite{jain_commentary_2013}, pure Ti, and pure C (graphene, graphite, diamond). These ensure that the ML potential reproduces the correct thermodynamic stability of the MXenes.

With an initial training database of structures, a first version of the ML potential was trained. After this, we gradually extended the training set by adding various defects and disordered MXene-like sheets by iterative training, which is a simple yet effective method for self-learning and gradually improving ML potentials. For defects we considered simple vacancies of both elements (and both inner and outer atomic layers for $n>1$ MXenes), self-interstitial atoms, adatoms in various positions, and antisites. A training iteration consists of creating a set of defected MXenes, relaxing them in short $NVT$ MD simulations using the current version of the ML potential, and adding them to the training database after obtaining the DFT energies and forces. Two iterations were enough to obtain a sufficiently diverse set of defect structures where further relaxations produced defects similar to existing structures. Disordered, molten, or half-recrystallised MXene-like sheets were also created in a similar iterative learning loop. Additionally, we manually created some defect saddle points from vacancy and adatom migration paths.

Finally, we observed in high-temperature MD simulations that a preliminary ML potential produced dense and energetically very stable unphysical clusters of C. These clearly artificial C clusters formed spontaneously when melting MXene sheets in MD. Again, we used iterative training to eliminate this artefact. After a few iterations, the ML potential learned from its mistakes and relaxation of previously predicted unphysical dense C clusters produced physically reasonable fullerene-like cages.

We believe this approach to construct the training database provides useful guidelines and insight that can be adopted when developing ML potentials for other MXenes, or when extending our database to include other elements or surface terminations.

Figure~\ref{fig:db} shows, apart from the training structures, also the train errors as root-mean-square errors (RMSEs) of the energies and force components with respect to DFT. The overall training errors for structures grouped as ``MXenes'' are 4.5 meV/atom and 0.19 eV/Å. For non-MXenes, the range of diversity and energy is much wider and since we do not require high accuracy for non-MXenes, these structures have lower weight in the training and consequently higher train errors. Overall, the training accuracy suggests good performance of the ML potential, although this must be further tested in dynamic simulations and calculation of real properties.

\subsection{Validation of the ML potential}

\begin{figure*}
    \centering
    \includegraphics[width=0.9\linewidth]{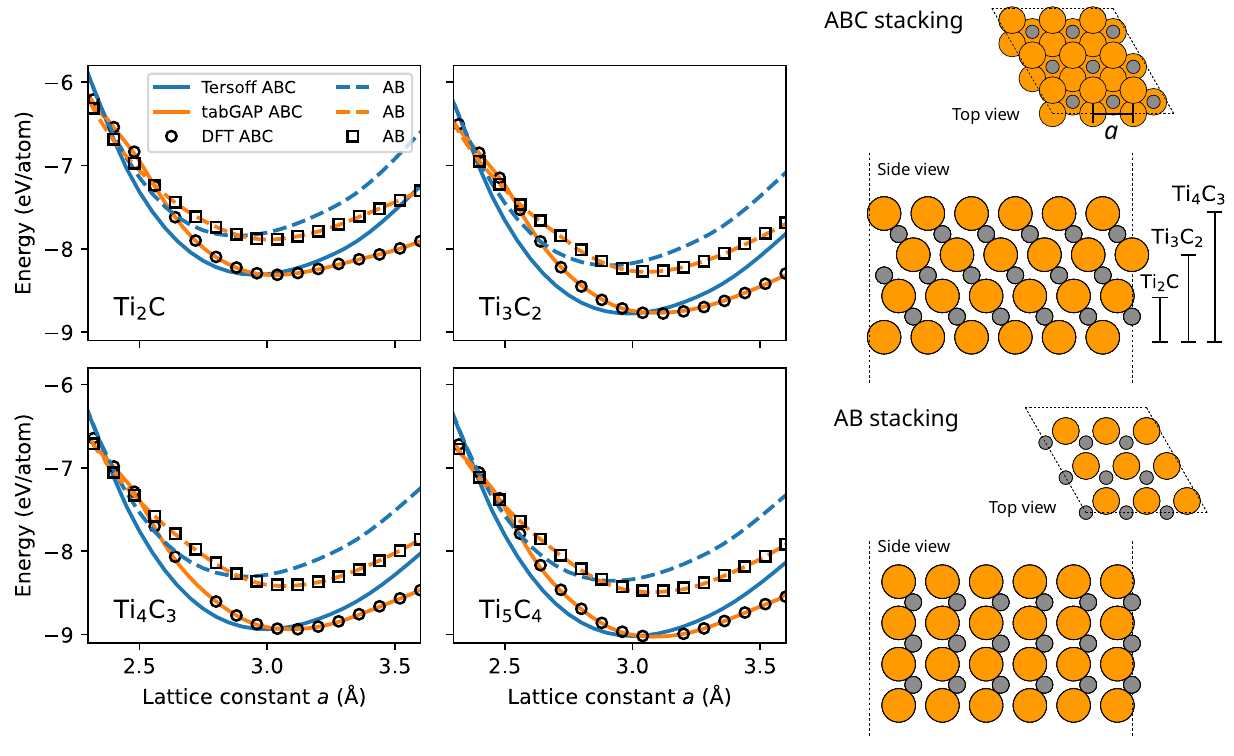}
    \caption{Energy as functions of the in-plane lattice constant $a$ for Ti$_{n+1}$C$_n$ MXene sheets with increasing $n$, compared between tabGAP, the Tersoff potential~\cite{plummer_bondorder_2022}, and DFT. The atom positions are fully relaxed at every lattice constant in all cases. Results are shown for both ABC and AB stackings, illustrated at the right. For visibility and direct comparison, the Tersoff data are shifted in energy to match the DFT energy minima for ABC stacking.}
    \label{fig:ev}
\end{figure*}

The first validation results for the ML potential are shown in Figure~\ref{fig:ev}, where the energy of Ti$_{n+1}$C$_n$ MXenes as functions of lattice constant are compared between the tabGAP, the Tersoff potential~\cite{plummer_bondorder_2022}, and DFT. Atom positions are fully optimised for every data point. Figure~\ref{fig:ev} shows that the ABC stacking is favoured for all thicknesses and methods and that the global minimum is shifted towards longer lattice parameters and lower energy as the MXene thickness increases. It is also clear that the tabGAP is quantitively more accurate than Tersoff, even though the latter also reproduces all trends qualitatively correctly.

\begin{table}
    \centering
    \caption{In-plane lattice constants ($a$) and monovacancy formation energies in Ti$_{n+1}$C$_n$ MXenes computed by DFT, tabGAP and the Tersoff potential. The DFT values in brackets are from spin-polarised DFT calculations.}
    \begin{tabular}{llccc}
    \toprule
        & & DFT & tabGAP & Tersoff \\
    \midrule
      Ti$_2$C & $a$ (Å) & 3.03 (3.08) & 3.03 & 2.95 \\
      & $V_\mathrm{Ti}$ (eV) & 7.67 (8.02) & 7.88 & 8.61 \\
      & $V_\mathrm{C}$ (eV) & 10.63 (10.85) & 10.50 & 11.41 \\
      \midrule
      Ti$_3$C$_2$ & $a$ (Å) & 3.10 & 3.09 & 2.98 \\
      & $V_\mathrm{Ti-top}$ (eV) & 8.01 & 7.79 & 8.37 \\
      & $V_\mathrm{Ti-mid}$ (eV) & 11.48 & 11.51 & 11.42 \\
      & $V_\mathrm{C}$ (eV) & 10.50 & 10.21 & 10.54 \\
      \midrule
      Ti$_4$C$_3$ & $a$ (Å) & 3.09 & 3.09 & 3.00 \\
      & $V_\mathrm{Ti-top}$ (eV) & 7.73 & 7.75 & 8.41 \\
      & $V_\mathrm{Ti-mid}$ (eV) & 12.16 & 11.64 & 11.26 \\
      & $V_\mathrm{C-top}$ (eV) & 10.31 & 10.13 & 10.57 \\
      & $V_\mathrm{C-mid}$ (eV) & 9.59 & 9.68 & 9.57 \\
    \bottomrule
    \end{tabular}
    \label{tab:props}
\end{table}

Table~\ref{tab:props} lists the in-plane lattice constants and monovacancy formation energies for all nonequivalent lattice sites of Ti$_{n+1}$C$_n$ up to $n=3$. Overall, the lattice constants and vacancy formation energies are in good agreement between DFT and tabGAP. The Tersoff potential reproduces the vacancy formation energies quite well but underestimates the lattice constants. Table~\ref{tab:props} also shows the effect of magnetism on the lattice constant and vacancy energetics for \ce{Ti2C} in brackets. The vacancy formation energies are different by only a few percent in spin-polarised DFT calculations.

\begin{figure*}
    \centering
    \includegraphics[width=\linewidth]{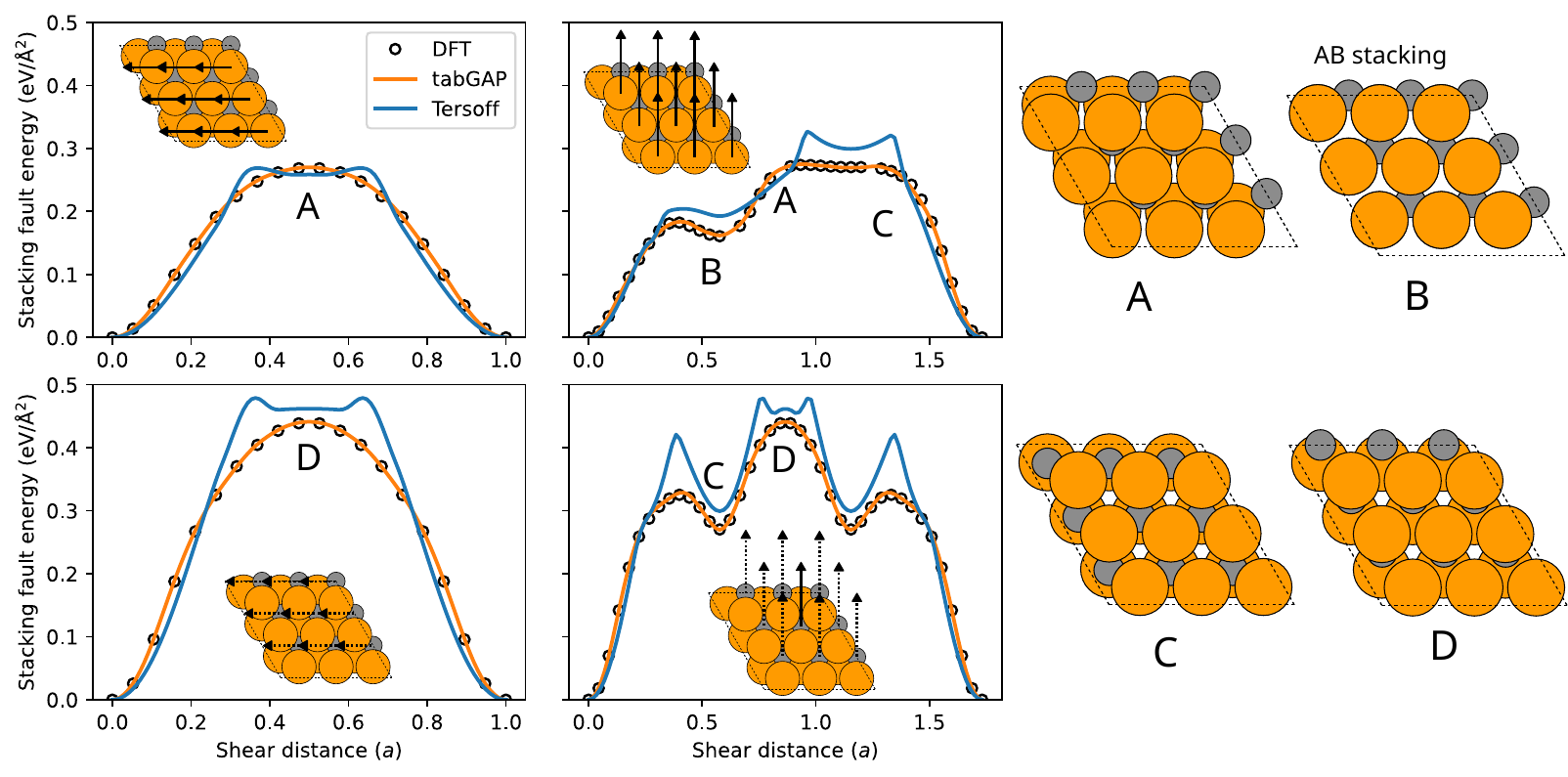}
    \caption{Theoretical shear strength of an ABC-stacked Ti$_2$C MXene sheet for collective shearing of an entire Ti monolayer (top) or C monolayer (bottom) along the two different illustrated directions ('armchair' and 'zigzag'). Atoms are relaxed along the surface normal direction only. Results are compared between tabGAP, the Tersoff potential, and DFT. The key saddle points and maxima are indicated and illustrated at the right, where for example the ideal AB stacking is reached when shearing a Ti layer in the zigzag direction 1/3 of a zigzag period.}
    \label{fig:shear}
\end{figure*}

In Figure~\ref{fig:shear}, energy profiles for various Ti and C monolayer shearing modes in Ti$_2$C are shown and compared between tabGAP, Tersoff, and DFT. The energies are given as stacking fault energies, i.e. energy differences divided by the monolayer (shear plane) area. Atoms are relaxed in the surface normal direction. These calculations serve as good validation tests of the ML potential and demonstrates the theoretical shear strength of the Ti$_2$C MXene. Figure~\ref{fig:shear} reveals that the tabGAP very accurately reproduces the DFT stacking fault profiles. The Tersoff potential is accurate for the magnitudes of the saddle points, but the profiles are not smooth and contain false local maxima and minima. We found that these are consequences of the short and relatively abrupt interaction cutoff, which is quite typical for Tersoff potentials and produces the unphysical features in Figure~\ref{fig:shear} when atoms enter and exit the interaction range.

\begin{figure*}
    \centering
    \includegraphics[width=0.7\linewidth]{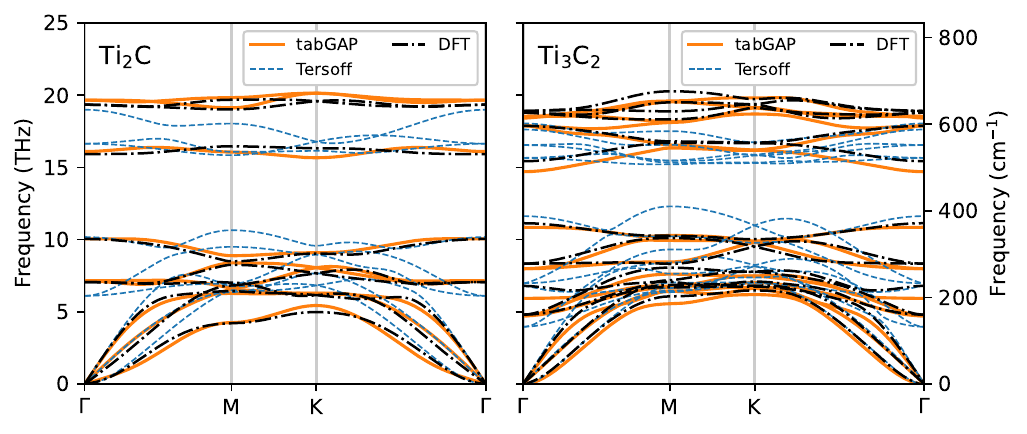}
    \caption{Phonon dispersion of \ce{Ti2C} and \ce{Ti3C2} MXenes compared between tabGAP, Tersoff, and DFT.}
    \label{fig:phon}
\end{figure*}

Figure~\ref{fig:phon} shows the phonon dispersion relations of \ce{Ti2C} and \ce{Ti3C2} MXenes computed with the tabGAP, Tersoff, and DFT. Overall, the tabGAP accurately reproduces the DFT phonon properties, implying that tabGAP predicts accurate near-equilibrium thermoelastic properties.

\begin{figure}
    \centering
    \includegraphics[width=\linewidth]{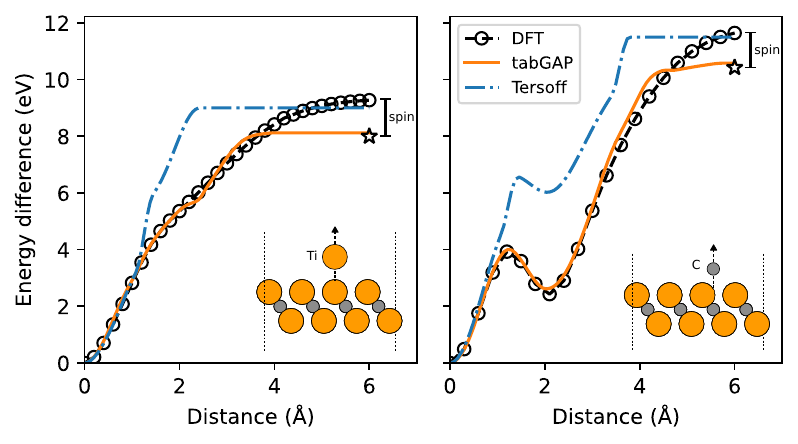}
    \caption{Potential energy landscape of a Ti atom (left) and C (right) atom dragged up along the surface normal in a rigid Ti$_2$C sheet, compared between DFT, tabGAP, and the Tersoff potential. The DFT calculations are done without spin polarization, which leads to incorrect energy as the dragged atom becomes isolated. The difference is marked by a line and the energy corrected by the difference between the nonmagnetic and spin-polarized isolated atom is shown with a star. The tabGAP is trained with the correct spin-polarized energy of the isolated atom and hence reproduces the correct limits indicated by the stars.}
    \label{fig:sput}
\end{figure}

As a test that directly probes the accuracy in defect formation due to irradiation and sputtering, in Figure~\ref{fig:sput} we show the energy difference for statically dragging a Ti or a C atom upwards from a Ti$_2$C sheet. The tabGAP follows the DFT data well, except that it approaches different limits as the dragged atom is detached from the surface and becomes isolated. The difference is due to incorrect isolated atom energy in the non-spin-polarized DFT used here, while tabGAP is trained with the correct spin-polarized isolated atom energies (although all other training structures are non-spin-polarized). Accounting for spin polarization of isolated atoms during sputtering has been shown to be important and hence the tabGAP results should be physically more correct than the non-spin-polarized DFT curves~\cite{kretschmer_threshold_2022}. The spin-corrected DFT limits are shown by stars in Figure~\ref{fig:sput} and coincide with the tabGAP limits. By definition, the limiting energy differences are the same as the vacancy formation energies in Table~\ref{tab:props}. When dragging a Ti atom, we note a slight deviation from the DFT curvature for tabGAP at around 2.4 Å in Figure~\ref{fig:sput}. Analysis revealed that this occurs when the Ti--C bonds are broken and is thus a direct consequence of the finite interaction range, however the overall trend is still smooth.

\begin{figure*}
    \centering
    \includegraphics[width=0.9\linewidth]{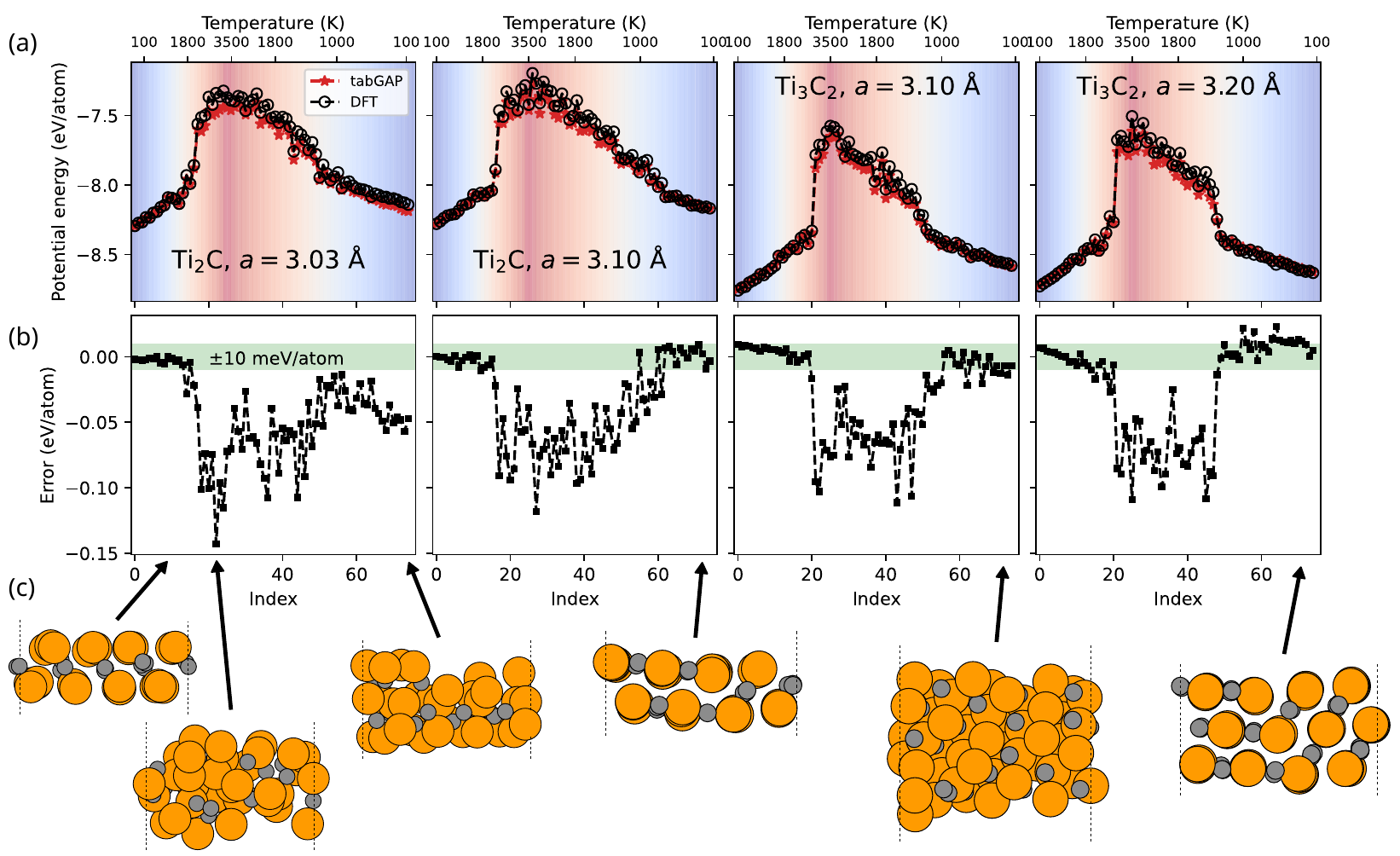}
    \caption{Melt-quench simulations of \ce{Ti2C} and \ce{Ti3C2} MXenes, each with two different initial lattice spacings. The simulations are done with tabGAP. Frames from the tabGAP trajectories are sampled and evaluated statically with DFT. The (a) panel shows the potential energy from tabGAP and DFT as a function of heating and cooling. The background colours visually indicate the heating and cooling stages. Panel (b) shows the energy error of the tabGAP compared to DFT, with a $\pm 10$ meV/atom error margin highlighted in green. The (c) panel shows representative snapshots from the trajectories.}
    \label{fig:melt-quench-valid}
\end{figure*}

Beyond static properties, we also ensured that the tabGAP is accurate, robust and numerically stable in high-temperature MD simulations. To also test the physical correctness, we performed melt-quench simulations by first heating up \ce{Ti2C} and \ce{Ti3C2} MXene sheets from 100 K to 3500 K over 500 ps, then cooling them down to 100 K for another 1 ns. We also considered two lattice spacings for each MXene. The cells were delibrately kept small (48 and 80 atoms), so that frames from the tabGAP MD trajectories could be extracted and computed in DFT. In this way the heating-cooling simulations provide a demanding test set of low-to-high temperature structures, both crystalline and disordered, for one-to-one comparison to DFT. The results are summarised in Fig.~\ref{fig:melt-quench-valid}, where the top panel shows the potential energy as function of trajectory index or temperature from tabGAP compared to DFT. The bottom panel shows the error in energy compared to DFT. 

During heating, the MXene sheets melt into disordered sheets with significantly higher potential energy, as shown in Fig.~\ref{fig:melt-quench-valid}. The increase of energy during melting (latent heat) is fairly consistent between tabGAP and DFT, although the error increases from less than 10 meV/atom for high-temperature crystalline MXene sheet to around 0.1 eV/atom for the molten sheets. During cooling, the disordered sheets partially recrystallise into mixed TiC- and MXene-like sheets. The appearence of TiC-like rocksalt symmetries is reasonable since bulk TiC is thermodynamically the most stable Ti--C phase and also corresponds to the structure of the MXene middle layers. The energy errors during recrystallisation drops to within 10 meV/atom for all cases except the first, where the quenched and recrystallised sheet is MXene-like but contains numerous defects and partial extra layers. Most importantly, however, the melt-quench tests show that the tabGAP is indeed stable and provides physically reasonable results even at extreme temperatures. To further confirm this, larger-scale melt-quench tests are also done with the tabGAP and reported in the Supplemental document.

Finally, large-scale simulations benefit from good computational speed. By design, this is a key strength of tabGAP, compared to many other ML potentials that may be orders of magnitude slower than traditional analytical interatomic potentials~\cite{zuo_performance_2020}. We benchmarked the speed of the tabGAP compared to the analytical Tersoff potential~\cite{plummer_bondorder_2022} and found that tabGAP is 20--40\% slower than Tersoff depending on MXene thickness (70--135 kiloatom-steps/s). This makes the tabGAP efficient enough for simulations of large length scales, long time scales, or extensive statistics. The speed was measured for MXenes of thickness $n=1$ to $n=3$ and planar size $10\times10$ unit cells in MD simulations with LAMMPS (version 10 Sep 2025) on a single CPU node (AMD Ryzen 7 Pro 5750G). The tests were done with the same standard LAMMPS compilation for both potentials and with default neighbor list settings, 10 K initial temperature, $NVE$ ensemble, and 200 time steps. The speed difference between tabGAP and Tersoff is due to the shorter interaction range of the Tersoff potential. This was confirmed by artificially extending the cutoff range of the Tersoff potential to that of tabGAP, after which tabGAP becomes 50--80\% faster than Tersoff.

\subsection{Irradiation and implantation simulations}

\begin{figure*}
    \centering
    \includegraphics[width=\linewidth]{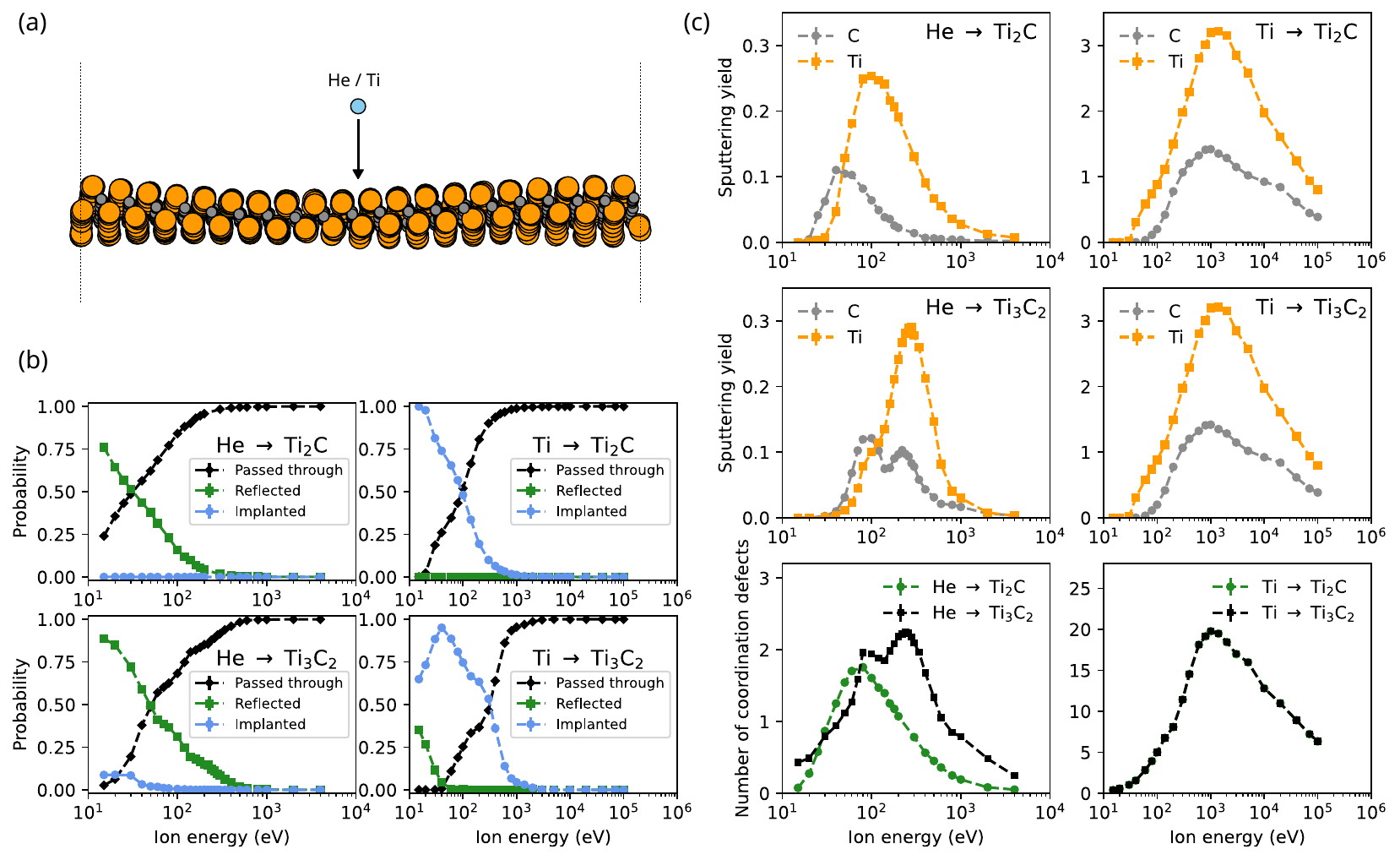}
    \caption{Main results from the irradiation simulations. The simulation setup (for Ti$_2$C) is illustrated in (a). (b) shows the probabilities of the incoming ion (He or Ti) passing through, reflecting, or being implanted in the MXene sheet as functions of ion energy for all four cases. (c) shows sputtering yields of C and Ti atoms in all four cases, as well as the number of coordination defects (counted as number of wrong-coordinated C atoms).}
    \label{fig:irrad}
\end{figure*}

We applied the ML potential in simulations of ion irradiation of single Ti$_2$C and Ti$_3$C$_2$ sheets to quantify the sputtering yields and understand the nature of defect production. The irradiation is simulated by incoming ions with varying kinetic energy, directed towards the MXene sheet normal to the surface as illustrated in Figure~\ref{fig:irrad}(a). To allow qualitative comparison to the experimental studies, we perform simulations with both light and heavier ions. For light ions we use He, as in the experiments by Benmoumen et al.~\cite{benmoumen_structural_2024}. For the heavier ions, Pazniak et al. used Mn, but for convenience we use Ti which has similar mass to Mn and already supported by the ML potential. The use of Ti is expected to be similar to Mn for ballistically controlled sputtering and damage creation, but cannot be assumed to be a substitute for Mn for more chemically controlled implantation and exact defect configurations. Note also that we use the word ``ion'' to be consistent with established terminology, even though we simulate atoms with no explicit charge state. We also neglect electronic stopping, since standard bulk methods are not necessarily accurate for 2D materials~\cite{liebsch_quantitative_2026}. However, these effects are most important at very high kinetic energies ($> 1$ MeV), while we here focus mainly on lower energies. He interactions are treated with repulsive pair potentials (see Methods). In total we performed around 1 million individual simulations to gather statistics for a range of incoming ion energies (10 000 ion impacts per energy). More details are provided in Methods.

The main results are summarised in Figure~\ref{fig:irrad}. Figure~\ref{fig:irrad}(b) shows the probabilities of the incoming ion passing through the MXene sheet, reflecting from the surface, or being implanted in or on the MXene as functions of incoming ion energy. Uncertainties are estimated from 95\% confidence intervals of binomial tests, although the error bars are smaller than the data points and not visible. The light and inert He ions are most likely to be reflected at low energies and passing through at higher energies. The crossover in probabilities increases with MXene thickness, but occurs between 10 and 100 eV for both Ti$_2$C and Ti$_3$C$_2$. The probability of an incoming He atom to be implanted into the MXene sheet is low but not zero, especially not for Ti$_3$C$_2$ at energies below 100 eV. For Ti impacts, implantation is the most likely result at energies below 100 eV for Ti$_2$C and below 300 eV for Ti$_3$C$_2$. Above 1 keV, most Ti ions pass through the MXenes, usually leaving some defects behind.

Figure~\ref{fig:irrad}(c) shows the calculated sputtering yields of C and Ti atoms as functions of ion energy. The error bars are standard errors of the mean, but again smaller than the data points. Both experimental studies reported preferential sputtering of Ti~\cite{pazniak_ion_2021, benmoumen_structural_2024}. Our simulations confirm that, except for very low He ion energies. It is also clear that Ti irradiation causes higher sputtering yields than He, by about one order of magnitude. Both of these results can be explained by the kinetic energy transfer during collisions. While Ti atoms sputter more easily than C because of the atomic arrangement, He ions transfer kinetic energy more efficiently to C atoms (due to smaller mass difference) which leads to higher C sputtering at low He ion energies. This effect is obscured once the He energies are high enough to cause recoils far above the sputter threshold for Ti atoms. For heavy-ion irradiation, the same effect promotes Ti sputtering at all energies.

There is a curious double-peak in the C sputtering yield trend for He irradiation of Ti$_3$C$_2$, while the trend for Ti sputtering has a single peak. One could expect this to be related to the two C layers and whether the C atoms sputter up or down from those layers, effectively creating different sputtering threshold energies. However, after visually analysing numerous simulation trajectories, we found no evidence for this. Instead, the double peak seems to be caused by a threshold energy at which the probability of sputtering \textit{both} C and Ti atoms from a single ion increases. At low energies, C sputters more often as explained above. As the energy increases, Ti sputtering becomes possible and rapidly increases, causing a drop in C sputtering. At around 150--200 eV there is enough energy that after preferentially sputtering a Ti atom, the He ion sometimes has energy to also sputter a C atom, producing a second peak. After the peak sputtering at 200--300 eV, both Ti and C sputtering yields decrease. At very high ion energies in all cases for both He and Ti irradiation, the sputtering yields approach zero due to decreasing cross sections for nuclear collisions and interactions~\cite{krasheninnikov_ion_2010}.

\begin{figure*}
    \centering
    \includegraphics[width=0.8\linewidth]{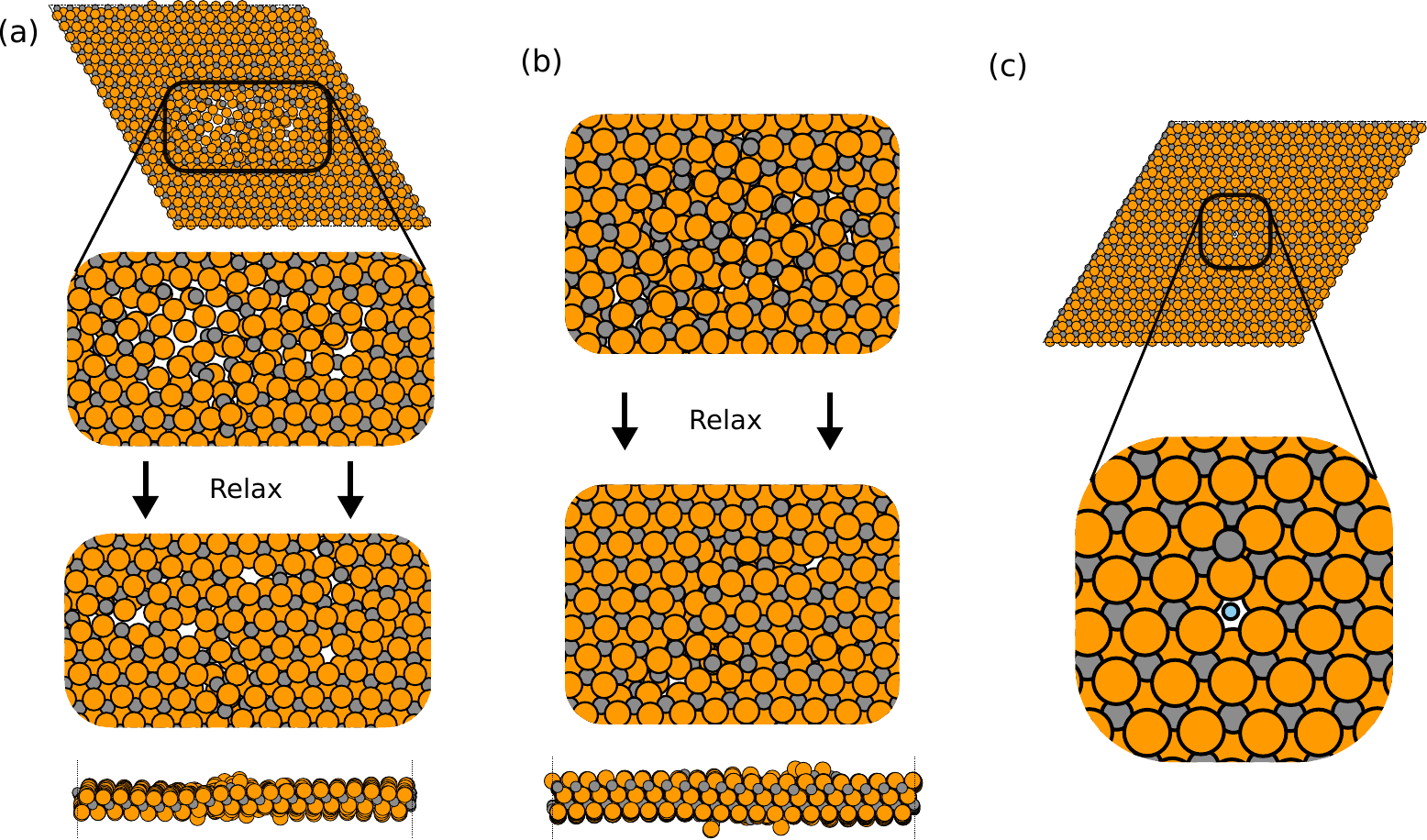}
    \caption{Snapshots from the irradiation simulations showing resilient stability and healing of the most extremely damaged MXene sheets, (a)-(b), and implantation of a He atom at low energy, (c). (a) shows a Ti$_2$C sheet 2 ps after impact of a 2 keV Ti atom, with the heavily damaged region zoomed-in. The bottom snapshots shows the same region after 100 ps relaxation (top and side views). (b) shows a similar case of efficient healing of a Ti$_3$C$_2$ sheet after a 2 keV Ti impact. (c) shows a rare He implantation event, where a low-energy (20 eV) He impact has displaced and replaced a C atom, which becomes an adatom on the bottom layer.}
    \label{fig:snap}
\end{figure*}

Figure~\ref{fig:irrad}(c) also shows the number of coordination defects in the MXene lattice as functions of ion energy, providing information about the damage created by the ion. It is clear that the thicker MXene absorbs more damage and that Ti ions cause more damage than He in terms of off-coordinated C atoms. For He irradiation, the thicker MXene shows a double peak similar to the C sputtering yield.

Heavy-ion irradiation causes significant damage, but we found that the MXene sheets still remain remarkably intact and show efficient self-healing and recovery. Hence, the coordination defects in Figure~\ref{fig:irrad}(c) should be considered upper limits, since the simulations are in the most highly damaged cases too short to allow full relaxation. To investigate the relaxation of the most heavily damaged MXene sheets, Figure~\ref{fig:snap}(a)-(b) shows snapshots from two representative cases of heavy damage and recovery after a 2 keV Ti ion impact in a Ti$_2$C and Ti$_3$C$_2$ sheet. The snapshots show significant disorder and damage created at 2 ps after the impact. After 100 ps of relaxation at 300 K, the lattice recovers efficiently into an intact sheet with vacancies left due to sputtering. The recovery is not perfect and leaves apart from the vacancies also adatoms and self-interstitial defects in the sheet, but most damage remains in the form of point-like defects. No open pores are ever observed for the ion energies considered, suggesting that MXene sheets are remarkably resilient and stable even during heavy irradiation. This is consistent with and explains the experimental observations in Ref.~\cite{pazniak_ion_2021} that the MXene sheets remain stable even after high-energy heavy-ion irradiation (60 keV Mn).

Finally, Figure~\ref{fig:snap}(c) shows one case of a successful He implantation into a single-layer Ti$_2$C sheet by displacing and replacing a C atom (which becomes an adatom). The probability for this is very low but not negligible (0.0001--0.0005 for He energies 15--40 eV) and for Ti$_3$C$_2$ sheets they are much higher (Figure~\ref{fig:sput}(b)). Since the simulations use purely repulsive screened Coulomb potentials for He interactions with Ti and C, it is unclear how reliable the implantation probabilities are. To investigate this further we computed the binding energy of a He atom inside the C vacancy in a \ce{Ti2C} sheet using DFT and the potentials. The results are 2.40 eV in DFT and 1.68 eV with the tabGAP and repulsive He pair potentials, using the isolated He atom as the reference for He. The two values are quite consistent considering that no refitting was done for the He potentials. The positive binding energies mean that the He-vacancy interaction is repulsive, and in fact that DFT predicts stronger repulsion than the potentials. From this one might conclude that the simulated implantation probabilities are overestimated, but that is not certain since the nonequilibrium implantation process does not necessarily correlate with the relaxed binding energy. Another aspect to consider is the trap strength of the implanted He atom, i.e. the migration energy for the He atom to escape the vacancy. As an estimate, we used the tabGAP with the He potentials to compute the migration barrier for escape, yielding 0.07 eV for the bare \ce{Ti2C} sheet. With such a low barrier, it is unlikely that the He atom remains implanted and stable for long times at room temperature. Nevertheless, the simulations indicate that ion implantation and in general defect engineering of MXenes by irradiation is a viable strategy, although it should be noted that all simulations here are on bare single MXene sheets. The effect of surface terminations on sputtering and implantation results requires further work and extension of the ML potential.

\section*{Conclusions}

The results have shown that robust, accurate, and fast ML interatomic potentials can be developed for MXenes, here for bare Ti$_{n+1}$C$_n$ sheets, opening up an avenue for machine-learning-driven atomistic simulations of the expanding chemical family of MXenes. Robustness is ensured by a diverse training database of structures covering thermoelastic, plastic, and defect properties. The training strategy developed here is repeatable and enables rapid development of ML potentials for other MXene compositions and as the foundation for including the relevant surface chemistry of MXenes. The ML potential was applied in a study of ion irradiation and implantation using both light (He) and heavy (Ti) ion impacts from 15 eV to 100 keV. The results quantified the extent and mechanisms of defect creation as well as ion implantation probabilities as functions of impact energy, providing insight for experimental defect engineering of Ti$_{n+1}$C$_n$ MXenes by ion irradiation. The simulations also quantified the sputtering yields of Ti and C atoms, confirming the preferential sputtering of Ti at higher impact energies. However, the onset of C sputtering occurs at lower energies than Ti sputtering for light-ion irradiation due to more efficient energy transfer.

\section*{Conflicts of interest}
There are no conflicts to declare.

\section*{Data availability}

The data supporting this article are available from the open repository \url{https://doi.org/10.5281/zenodo.21234116}.

\section*{Acknowledgements}

Funding from the Research council of Finland through the OCRAMLIP project, grant number 354234 is acknowledged.
Grants of computer capacity from CSC - IT Center for Science are gratefully acknowledged.





\bibliography{mybib} 
\bibliographystyle{rsc} 
\end{document}


\title{Supplemental document of: Machine-learned interatomic potential for titanium carbide \mbox{MXenes:} Application to ion irradiation simulations}

\author{Jesper Byggmästar}
\email{jesper.byggmastar@helsinki.fi}
\affiliation{Department of Physics, P.O. Box 43, FI-00014 University of Helsinki, Finland}

\maketitle

\tableofcontents

\clearpage

\section{Cross-validation results}

\begin{figure}[h]
    \centering
    \includegraphics[width=\linewidth]{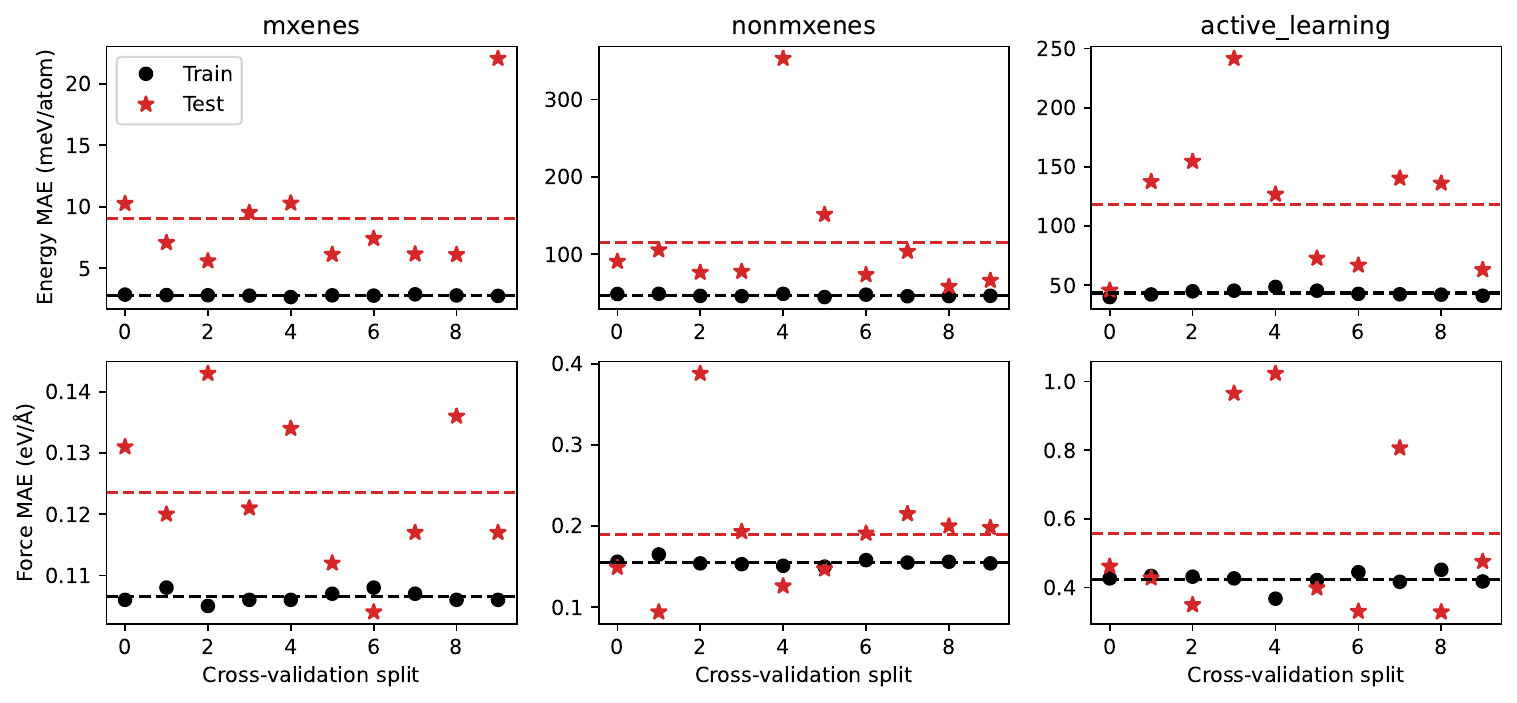}
    \caption{Train and test energy and force errors from $k$-fold cross-validation with $k=10$. The splits are done randomly within three groups of structures: MXenes ($N=774$), non-MXenes ($N=679$), and active/iterative learning ($N=44$). The averages errors are indicated by dashed lines.}
    \label{fig:cross}
\end{figure}

Figure~\ref{fig:cross} shows results from $k$-fold cross-validation of the training data, where the full training database is split into $k=10$ parts and training is done so that each of the 10 splits is used as test data when training on the remaining 9 splits. The database is separated into three groups; MXenes, non-MXenes, and structures from active/iterative learning (see main text), and each group is split randomly into 10 parts. 

Figure~\ref{fig:cross} reveals that the energies are somewhat overfitted, although still with reasonable test errors, while forces are more consistent between train and test. More alarmingly, Fig.~\ref{fig:cross} shows clear outliers where the test errors are dramatically higher than the train error. Further inspection reassured that the outliers are a result of the training database construction, where the outlier test errors come from structures that are quite extreme (in energy) and unique, i.e. representing the 'edges' of the configuration space that are never reached in practical simulations and where high accuracy is not needed. Specifically, in the 'MXenes' group, the high test energy error outlier for split 9 in Fig.~\ref{fig:cross} comes from the fact that the MXene unit cell with the strongest compressive strain is in the test set. This structure has en energy around 3 eV/atom higher than the closest structure (and about 8.5 eV/atom higher than the ground state), meaning that it is a result of the ML potential extrapolating to unphysically dense MXenes. Similarly, the energy outlier in the 'Non-MXenes' group comes from when the densest pure-carbon FCC unit cell (with an energy of $+55$ eV/atom) is in the test split. More generally, the significantly higher test than train errors are always affected by similar but not always so extreme structural outliers (i.e. first or last point in a strain or volume interval). The active learning errors are even more spread out, but this groups only has 4--5 structures in each test split and more importantly all structures are by design important to include in the training.

From the cross-validation results and the above analysis we conclude that with the strategy for training data generation used here, i.e. relying heavily on sampling strain or volume intervals as well as iteratively generating new structures and actively fixing issues by sampling structures from simulations with preliminary potential versions, randomly splitting the database into test and train sets is dangerous. Additionally, the cross-validation results are not necessarily representative of the model accuracy due to the outliers. Although outlier detection is an important aspect of cross-validation, the outliers detected here are the most extreme but already known structures of the training database.

\section{Melt-quench validation versus Tersoff}

\begin{figure}[h]
    \centering
    \includegraphics[width=\linewidth]{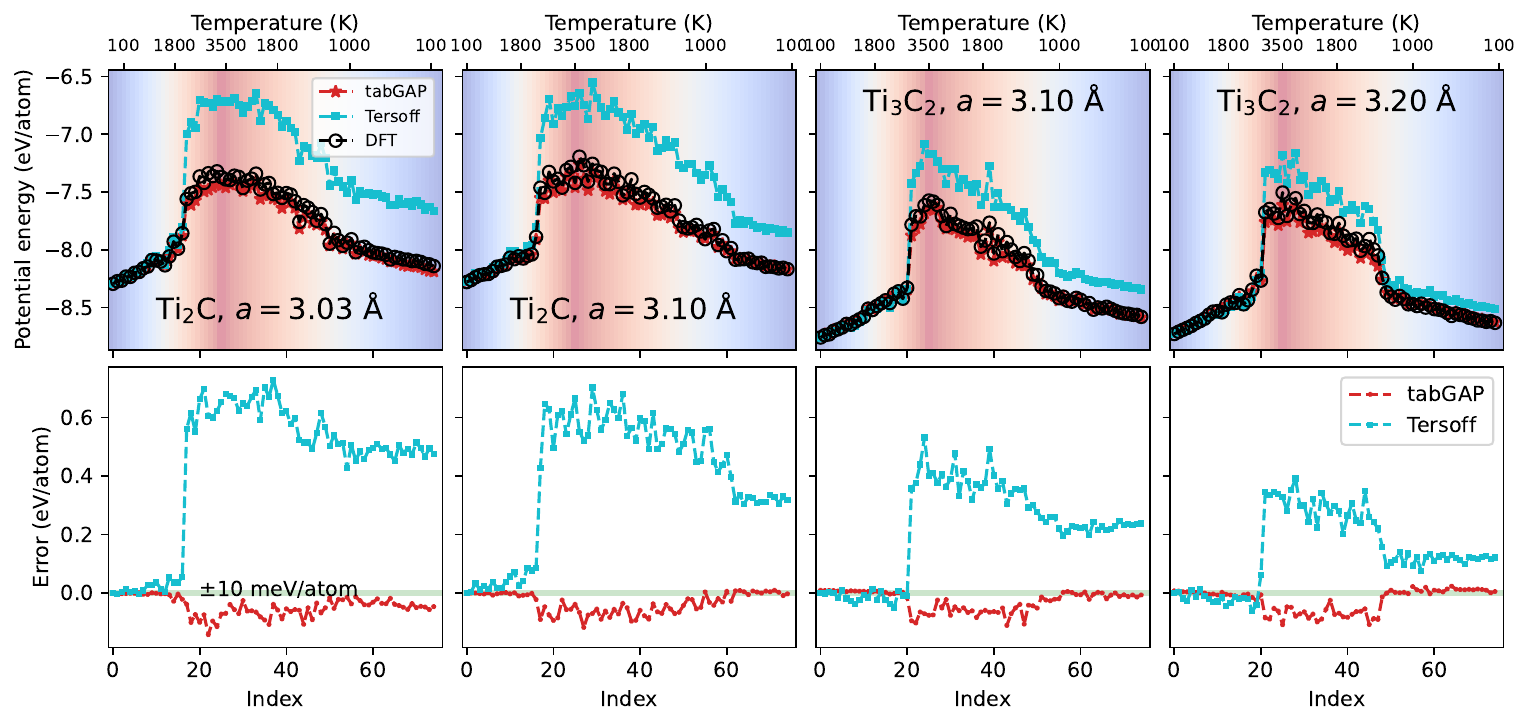}
    \caption{Melt-quench simulation test shown in the main text but including results from the Tersoff potential.}
    \label{fig:heatup_ters}
\end{figure}

Fig.~\ref{fig:heatup_ters} shows the melt-quench test discussed in the main text where, in addition to DFT, energies given by the Tersoff potential for each tabGAP-simulated frame are shown. Note that the error scale is different compared to the main text due to higher errors in Tersoff, but the same $\pm10$ meV/atom interval is highlighted in green for scale.

\section{Larger-scale melt-quench simulations}

\begin{figure}[h]
    \centering
    \includegraphics[width=0.8\linewidth]{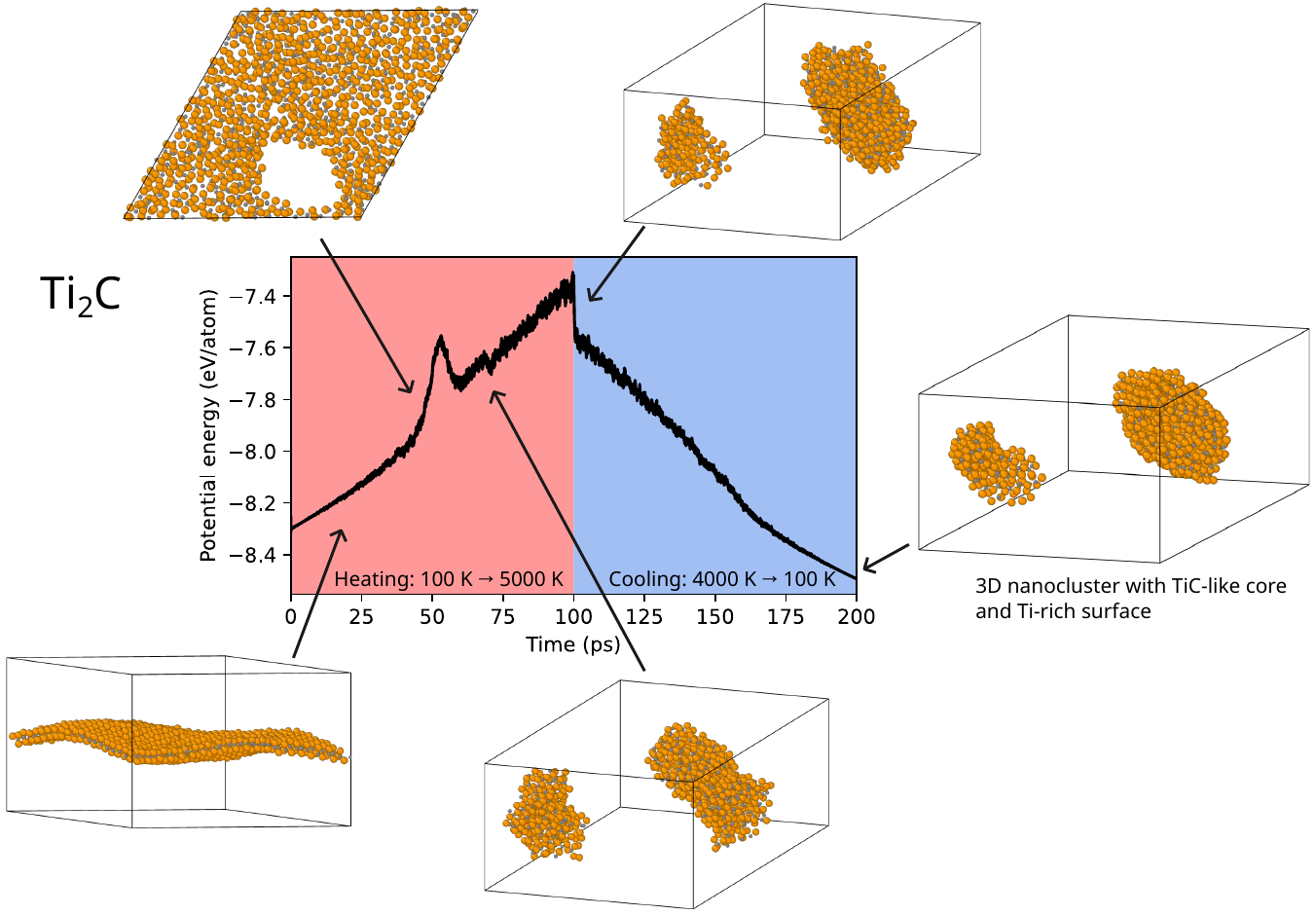}
    \caption{Snapshots from the heating-cooling simulations of the \ce{Ti2C} MXene described in the main text. Note that periodic boundary conditions are used, so that the molten and recrystallised nanocluster is connected and not two separate clusters as it appears.}
    \label{fig:heatn1}
\end{figure}

\begin{figure}[h]
    \centering
    \includegraphics[width=0.8\linewidth]{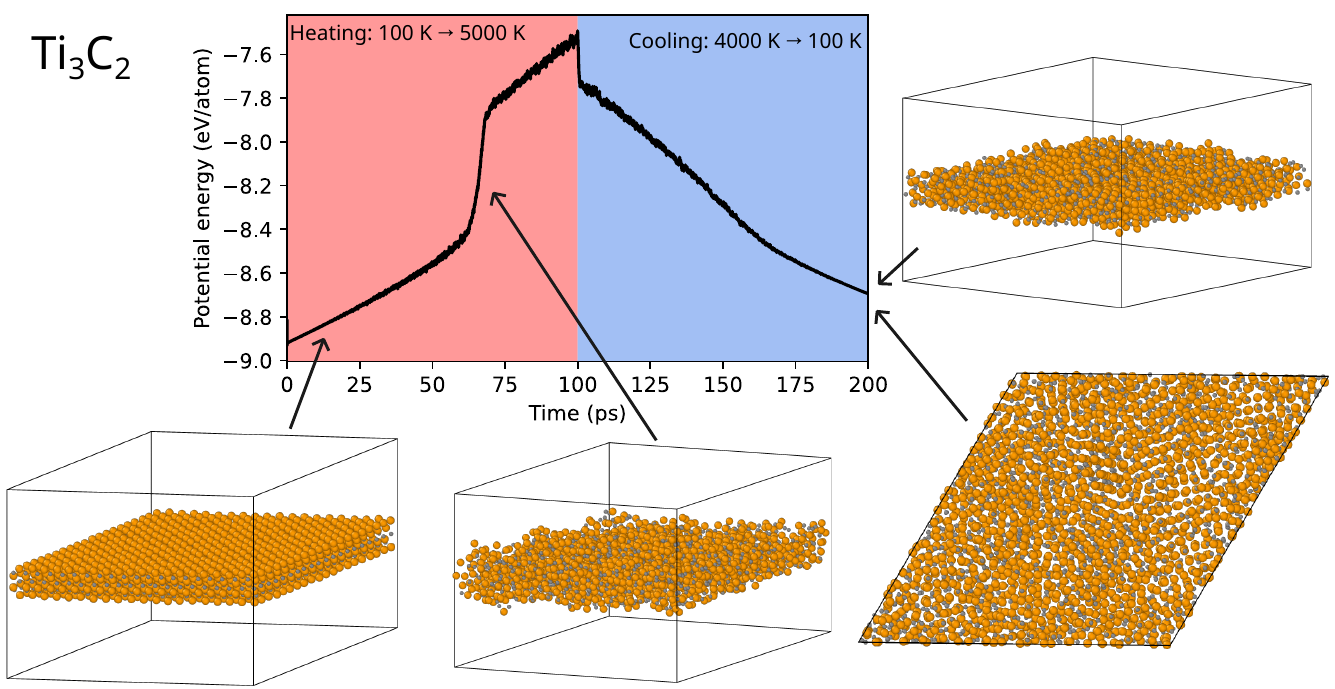}
    \caption{Snapshots from the heating-cooling simulations of the \ce{Ti3C2} MXene described in the main text.}
    \label{fig:heatn3}
\end{figure}

Figure~\ref{fig:heatn1} shows snapshots from larger-scale heating-cooling simulations of the \ce{Ti2C} MXene and Figure~\ref{fig:heatn3} from \ce{Ti3C2} MXene. The cells are too large to validate against DFT and mainly done to test the physical correctness of the tabGAP. We performed melt-quench simulations by first heating up the MXene sheets from 100 K to 5000 K over 100 ps, then cooling them down to 100 K for another 100 ps. At 2300--2500 K, the \ce{Ti2C} sheet rips apart and melts into a three-dimensional nanocluster. During cooling, the core of the nanocluster partially recrystallises into the rocksalt TiC phase with the excess Ti atoms covering the surface. This is a reasonable prediction by the tabGAP, since bulk TiC is thermodynamically the most stable Ti--C phase and also corresponds to the structure of the MXene middle layers. The thermodynamically more stable \ce{Ti3C2} sheet reaches 3200 K before melting but remains as a stable 2D sheet. During cooling it partially recrystallises to MXene-like structure but with residual C--C bonds (bond lengths $\sim$1.4 Å like in graphene and graphite). The energy of the melt-quenched sheet is significantly higher (0.24 eV/atom) than the perfect \ce{Ti3C2} MXene, which means that further MXene recrystallisation and defect annealing would be energetically favourable.

\clearpage
\section{Effect of magnetism}

\begin{figure}[h]
    \centering
    \includegraphics[width=0.5\linewidth]{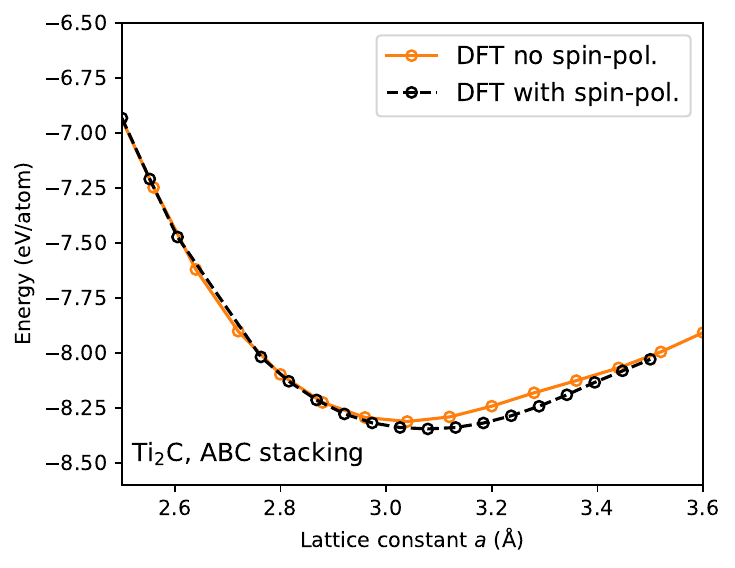}
    \caption{Energy versus lattice constant for a MXene with and without spin polarisation in DFT.}
    \label{fig:mag}
\end{figure}

Figure~\ref{fig:mag} shows the energy versus lattice constant for the single-layer \ce{Ti2C} MXene sheet from DFT calculations with and without spin-polarisation. Atom positions are relaxed. Spin-polarisation produces a magnetic state close the equilibrium that reduces the minimum energy by 33 meV/atom and increases the relaxed lattice constant from 3.03 Å to 3.08 Å. For compressed MXenes, the relaxed magnetic moments vanish and the energy follows the non-spin-polarised trend. Convergence issues become prevalent and highly strained lattices and spin-polarised calculations, evident from a few missing data points at 2.6--2.8 Å in Figure~\ref{fig:mag}. This is also the primary reason for neglecting spin-polarisation in the training data of the machine-learned potential.

\clearpage
\section{Repulsive potentials}

\begin{figure}[h]
    \centering
    \includegraphics[width=\linewidth]{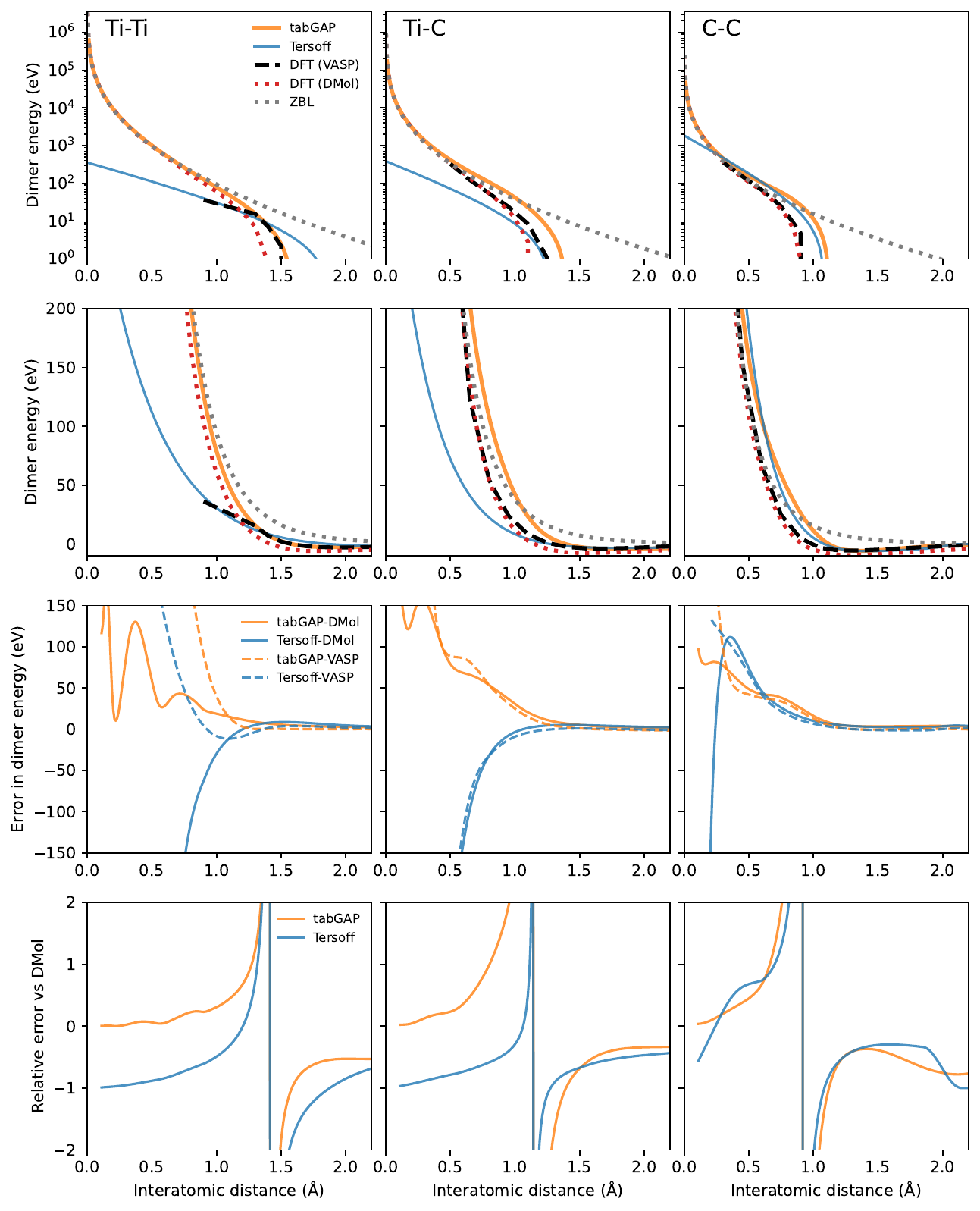}
    \caption{Repulsive parts of dimer curves for all pairs compared between tabGAP, Tersoff, DFT (VASP), all-electron DFT DMol, and the universal ZBL potential. The two bottom rows show the errors (direct and relative) of tabGAP and Tersoff compared to VASP and DMol.}
    \label{fig:rep}
\end{figure}

Figure~\ref{fig:rep} shows the repulsive parts of all dimer curves for tabGAP and Tersoff as well as the true and relative errors compared to DFT (VASP and all-electron DMol data~\cite{nordlund_repulsive_2025a}). The figure illustrates the physically correct screened Coulomb repulsion included in tabGAP (see main text). The Tersoff potential does not include screened Coulomb repulsion, but is still reasonably accurate down to 1 Å for Ti-Ti and Ti-C and down to 0.5 Å for C-C pairs. The tabGAP overshoots the Ti-C and C-C repulsion at intermediate distances ($>0.7$ Å) but smoothly approaches and joins the baseline screened Coulomb repulsion at shorter distances.

\begin{figure}[h]
    \centering
    \includegraphics[width=0.8\linewidth]{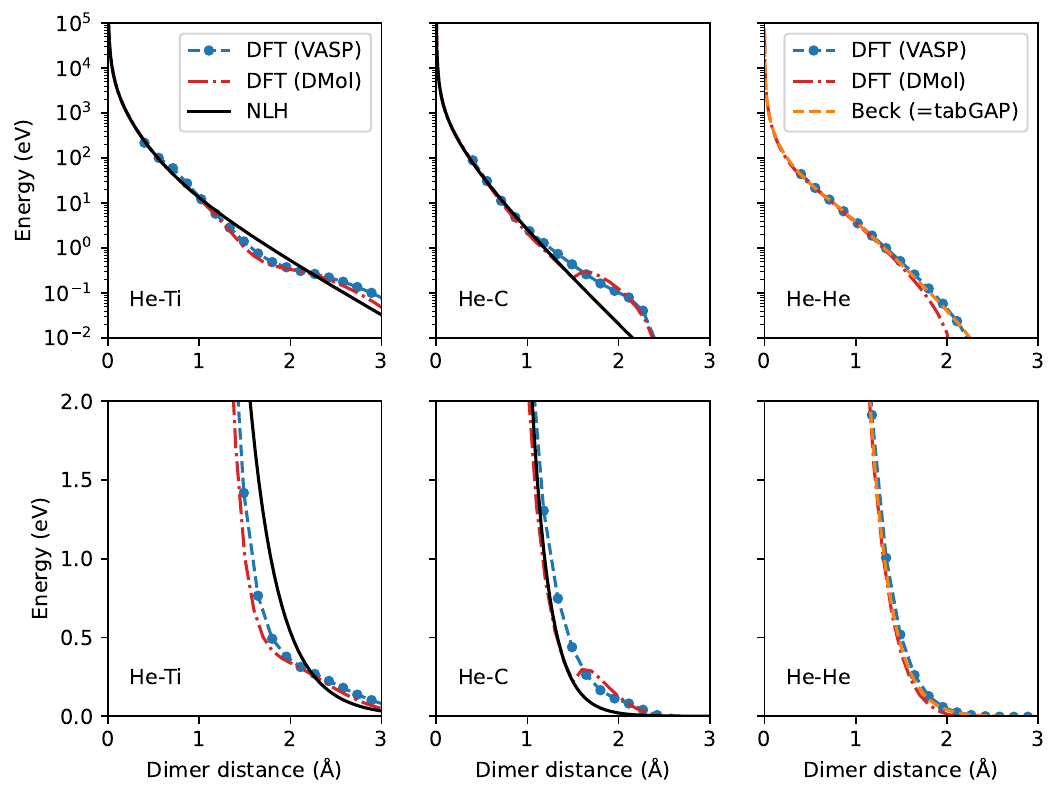}
    \caption{Repulsive parts of the He dimer curves compared between DFT and the NLH and Beck potentials used in the He irradiation simulations with tabGAP.}
    \label{fig:repHe}
\end{figure}

Figure~\ref{fig:repHe} shows the repulsive parts of the He dimer curves for DFT (VASP and DMol) compared to the NLH potentials~\cite{nordlund_repulsive_2025a} and He-He Beck potential~\cite{beck_new_1968} used in the He irradiation simulations with tabGAP.

\clearpage
\section{Minimum distances in irradiation simulations}

\begin{figure}[h]
    \centering
    \includegraphics[width=0.8\linewidth]{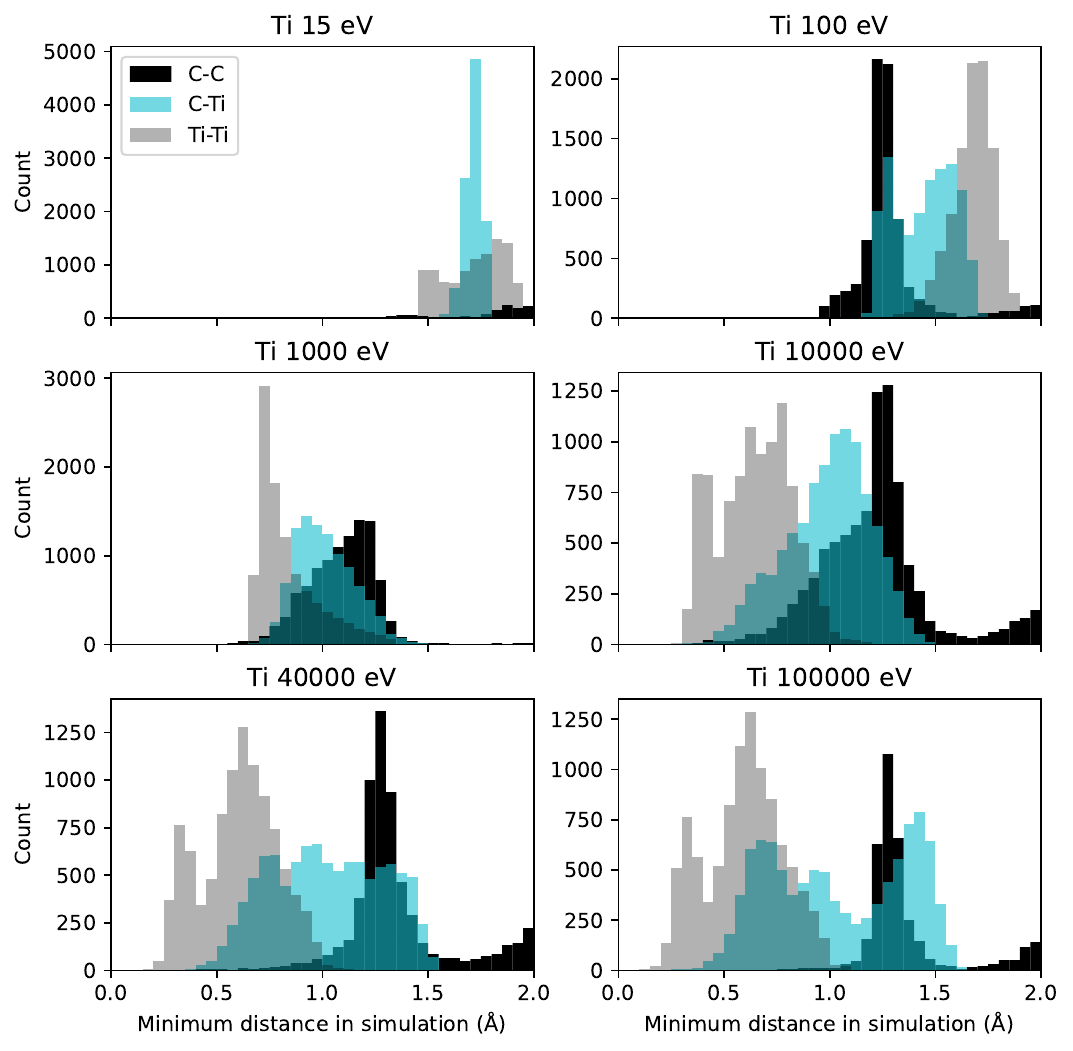}
    \caption{Histograms of minimum distances encountered in the Ti ion irradiation simulations.}
    \label{fig:rmin_hist_Ti}
\end{figure}

\begin{figure}[h]
    \centering
    \includegraphics[width=0.8\linewidth]{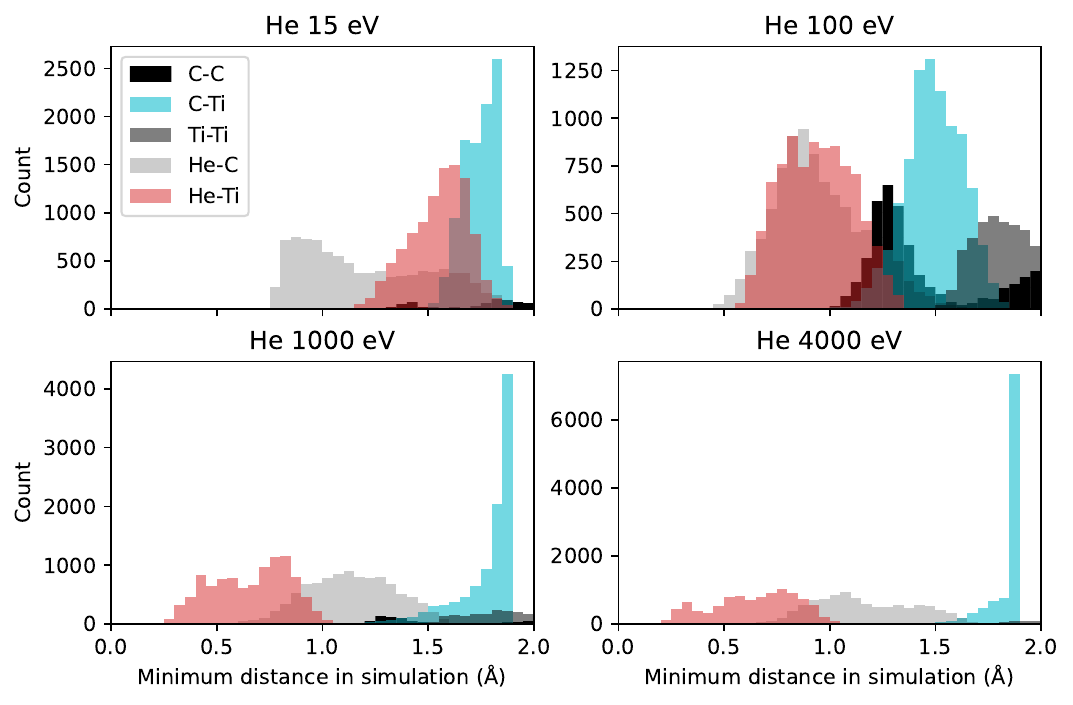}
    \caption{Histograms of minimum distances encountered in the He ion irradiation simulations.}
    \label{fig:rmin_hist_He}
\end{figure}

To investigate the importance of the repulsive potentials, Figures~\ref{fig:rmin_hist_Ti} and~\ref{fig:rmin_hist_He} show histograms of the minimum interatomic distances encountered in the irradiation simulations for a few selected ion energies. Only distances below 2 Å are considered and shown. From Figure~\ref{fig:rmin_hist_Ti} it is clear that already at 1 keV Ti ion irradiation, a large fraction of the simulations include collisions with distances 0.5--1.0 Å, where the Coulombic pair repulsion dominates and is crucial. For He irradiation, where the energy transfer to Ti is less efficient, He-Ti and He-C pairs reach short distances while Ti-Ti and Ti-C pairs are rarely close.

\begin{figure}[h]
    \centering
    \includegraphics[width=0.9\linewidth]{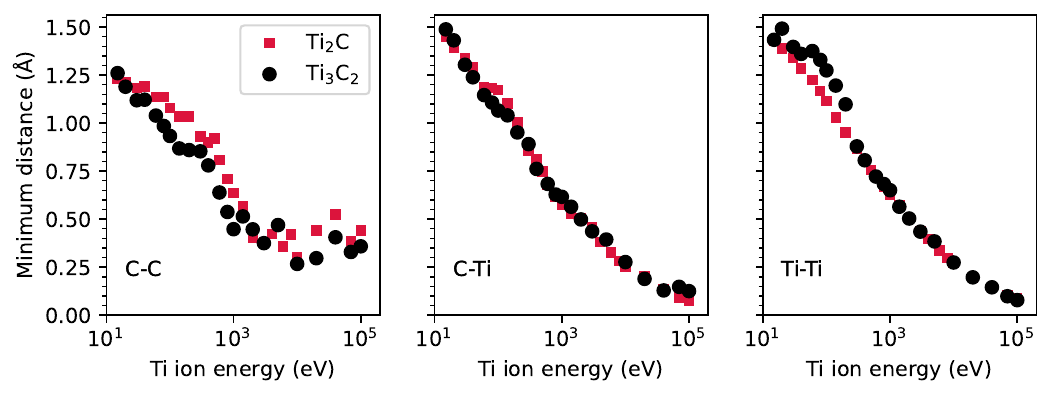}
    \caption{Minimum distance in all 10 000 simulations for each Ti ion energy.}
    \label{fig:rmin_min_Ti}
\end{figure}

\begin{figure}[h]
    \centering
    \includegraphics[width=0.9\linewidth]{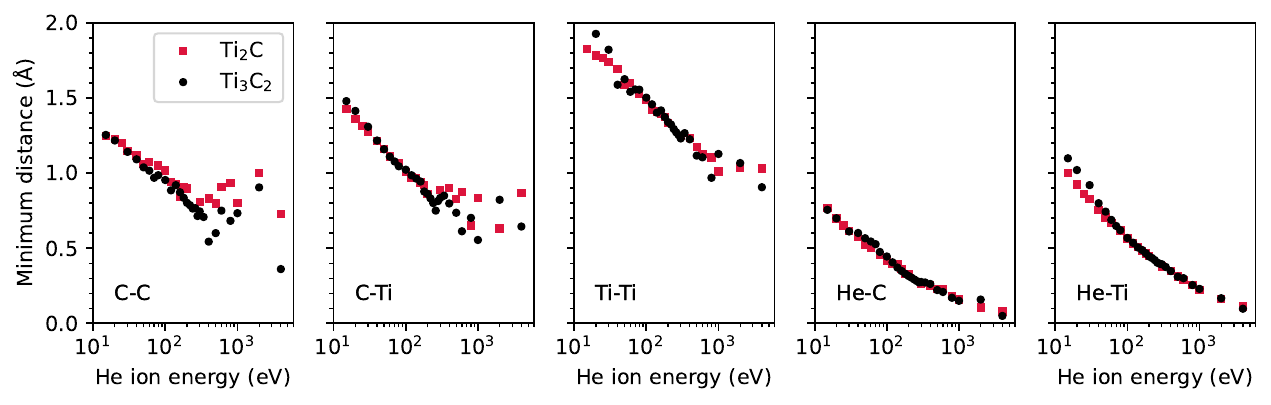}
    \caption{Minimum distance in all 10 000 simulations for each He ion energy.}
    \label{fig:rmin_min_He}
\end{figure}

For clarity, Figures~\ref{fig:rmin_min_Ti} and \ref{fig:rmin_min_He} show the minimum (across all 10 000 simulations per energy) of the minimum distances.

\clearpage
\bibliography{mybib}